# A 1D Model of Liquid Laminar Flows with Large Reynolds Numbers in Tapered Microchannels

Leonid Pekker

FujiFilm Dimatix Inc., Lebanon NH 03766 USA

**Abstract**

In this article, we construct a novel 1D-model of microfluidic laminar flows in tapered circular and rectangular channels assuming the flow in channels fully developed. In the model, we take into account the inertance and dynamic pressure terms. The model can be used for a wide range of flows: from the pure capillary flow regime, where the capillary forces are the main driver of the liquid in the channel, to the external pressure flow regime where the external pressure applied to the liquid at the entrance to the channel is much larger than the capillary pressure in the channel, so that the capillary force can be ignored. We apply the model to rectangular Y-shape junctions, where the base channel is connected to a reservoir and the end channels are exposed to atmospheric air. We show that, in asymmetric Y-shape junctions, there can be a time of meniscus arrest, where only one of the two channels with a smaller radius fills, and, the other one, with a larger radius, is arrested. The time of meniscus arrest decreases with an increase in the applied external pressure; when this pressure becomes large enough, the meniscus arrest disappears. However, if the ratio of the radii of the end channels is large, the meniscus arrest does not appear because the flow resistance in the channel with the smaller radius is much large than in the channel with the larger radius so that the initial flow in the wider channel cannot be stopped by the differences in the surface tension pressures in these channels. In this article, we also investigate the applicability of the fully developed flow approximation assumed in the model.

## 1. Introduction

The movement of liquids through capillary channels, due to the wetting of the liquid on the wall, was described more than a century ago [1-3]. In these works, the Reynolds numbers are assumed to be small so that the inertance and dynamic pressure can be ignored, and the flows are assumed to be fully developed. In the case of a straight capillary tube, this leads to "Washburn's law", which states the position of the meniscus in the capillary is proportional of square root of the capillary filling time. This law is widely used to describe the dynamics of imbibition in porous media [4-6].

Numerous papers were published to describe the movement of the liquid in different shapes of capillary channels consisting of uniform cross-sectional segments connected to each-other [7-14]. A model of imbibition process in axisymmetric non-uniform cross-sectional capillary channels was constructed in [15], and models of the inverse problem of capillary filling were constructed in [16, 17].

In [18], the authors consider the process of filling a system of three capillary channels connected by a Y-shape junction, Fig. 1. In their model, the channels are assumed to be rectangular with constant cross-sections (not tapered); and the junction filling is assumed to be instantaneous. This model has been used to calculate flows in porous media [18-20].

In all these works, [1-20], the effect of inertia and the dynamic pressure is neglected, $Re \ll 1$, so that the process of filling the capillary is governed by balance of the viscosity drag, the capillary force, and external force applied to the liquid.

A classical model of the flow of liquids in a straight circular capillary channel (not tapered) taking into account the inertia and dynamic pressure was suggested in [21]. In this work, the governing equation was derived from energy arguments and the flow was assumed to be fully developed. In [22-27], model [21] and its modifications taking into account a possible deviation from the fully developed flow approximation were used to investigate the inertia effect. In all these works, [21-27], the capillary channels are assumed to be circular with a constant cross-section.

In this article, we construct a 1D model of liquid laminar flow in tapered circular and rectangular microchannels based on the incompressible Navier-Stokes equations. In this model, we take into account the inertia and the dynamic pressure of the liquid and assume that the flows in the channels are fully developed and that the gravitational forces are small and can be neglected. Then, we apply this model to simulate flows in a three-channel Y-shape capillary network where the base channel is connected to a pressurized reservoir and the end channels are exposed to the ambient atmospheric air. A principal schema of our three-channel Y-shape capillary network model is shown in Fig. 2. To model a real Y-shape capillary network junction we have to convert the geometry of a real network to an equivalent 1D model, shown in Fig. 2.

We also construct a reduced microchannel model in which we drop the inertance and dynamic pressure terms, resulting in creeping flow microchannel regime model for tapered channels. In the case of three-channel Y-shape capillary network with not-tapered rectangular channels, this model reduces to model [18].

We show that both models, full and reduced, produce the same results when the external pressure applied to the liquid at the entrance to the network, $P_0(t)$ in Fig. 2, is set to zero (capillary imbibition process), the case of $Re \ll 1$. With an increase in the external pressure, the models begin diverging from each other. We also show that with an increase in the external pressure, the duration of meniscus arrest time in asymmetric Y-shape junction decreases; when this pressure becomes large enough, the meniscus arrest disappears. If the ratio of the entrance cross-sections of the end channels is large, Fig, 2, the meniscus arrest does not appear.

At the reservoir-channel interface, Fig. 2, the flow at the entrance to the channel is usually assumed to be plugged so that the flow becomes fully developed only beyond some transition distance in the channel. Using model [28], we derive the length of this transition region and compare it with the channel length to verify if our model assumption of the fully developed flow approximation is reasonable or not.

The paper is organized as follows. In Section 2, we derive a 1D-flow equation for circular tapered channels and, in Section 3, for rectangular tapered channels. Then, in Sections 4 and 5, based on the flow equations obtained in Sections 2 and 3, we construct the models, full and reduced, describing the filling of three-channel Y-shape capillary networks consisting, correspondingly, of circular and rectangular channels. In Section 6, we introduce a model for the length of the transition region in which a flow transforms from the plug-shape flow to the fully-developed-shape flow. The numerical results are presented in Sections 7 and the discussion in Section 8.

## 2. 1D-model of flow in a tapered circular channel

In a tapered circular channel, Fig. 3, a set of equations describing axisymmetric flow of an incompressible fluid can be written as

$$\frac{\partial v_z}{\partial t} + v_r \frac{\partial v_z}{\partial r} + v_z \frac{\partial v_z}{\partial z} = -\frac{1}{\rho} \cdot \frac{\partial P}{\partial z} + \frac{\mu}{\rho} \left( \frac{1}{r} \frac{\partial}{\partial r} \left( r \frac{\partial v_z}{\partial r} \right) + \frac{\partial^2 v_z}{\partial z^2} \right) \quad (2.1)$$

$$\frac{\partial v_r}{\partial t} + v_r \frac{\partial v_r}{\partial r} + v_z \frac{\partial v_r}{\partial z} = \frac{\mu}{\rho} \left( \frac{1}{r} \frac{\partial}{\partial r} \left( r \frac{\partial v_r}{\partial r} \right) - \frac{v_r}{r^2} + \frac{\partial^2 v_r}{\partial z^2} \right) \quad (2.2)$$

$$\frac{\partial v_z}{\partial z} + \frac{1}{r} \frac{\partial}{\partial r} (r v_r) = 0 \quad (2.3)$$

where Eqs. (2.1) and (2.2) are the Navier-Stokes equations correspondingly for $z$-momentum and $r$-momentum; Eq. (2.3) is the volume conservation equation; $\rho$ and $\mu$ are the mass density and the viscosity of the liquid, respectively; and $P$ is the pressure. In the model, we assume the non-slip boundary conditions at the surface of channel:

$$(v_z)_R = 0 \quad \text{and} \quad (v_r)_R = 0 \quad (2.4)$$

where $R(z)$ is the radius of the channel as a function of the $z$-coordinate; and Z is the length of the jet as shown in Fig. 3.

Integrating Eq. (2.1) over the cross-section of the channel and taking into account Eqs. (2.3) and (2.4), we obtain

$$\int_0^R \left( \frac{\partial v_z}{\partial t} + v_r \frac{\partial v_z}{\partial r} + v_z \frac{\partial v_z}{\partial z} \right) r dr = \int_0^R \left( -\frac{1}{\rho} \cdot \frac{\partial P}{\partial z} + \frac{\mu}{\rho} \left( \frac{1}{r} \frac{\partial}{\partial r} \left( r \frac{\partial v_z}{\partial r} \right) + \frac{\partial^2 v_z}{\partial z^2} \right) \right) r dr \quad (2.5)$$

LHS: $\frac{\partial}{\partial t} \left( \int_0^R v_z r dr \right) + \int_0^R \left( \frac{1}{r} \frac{\partial (v_z v_r r)}{\partial r} - \frac{v_z}{r} \frac{\partial (v_r r)}{\partial r} + v_z \frac{\partial v_z}{\partial z} \right) r dr =$

$= \frac{\partial}{\partial t} \left( \int_0^R v_z r dr \right) + \int_0^R \left( \frac{\partial (v_z v_r r)}{\partial r} + r v_z \frac{\partial v_z}{\partial z} + r v_z \frac{\partial v_z}{\partial z} \right) dr =$

$= \frac{\partial}{\partial t} \left( \int_0^R v_z r dr \right) + \int_0^R \left( \frac{\partial (v_z v_r r)}{\partial r} + \frac{\partial (r v_z^2)}{\partial z} \right) dr =$

$$= \frac{\partial}{\partial t}\left(\int_0^R v_z r\, dr\right) + (v_z v_r r)_{r=0}^{r=R} + \frac{\partial}{\partial z}\int_0^R (rv_z^2\, dr) - (rv_z^2)_{r=R}\frac{dR}{dz} =$$

$$= \frac{\partial}{\partial t}\left(\int_0^R v_z r\, dr\right) + \frac{\partial}{\partial z}\int_0^R (rv_z^2\, dr) \tag{2.6}$$

RHS: $-\frac{R^2}{2\rho}\cdot\frac{\partial P}{\partial z} + \frac{\mu}{\rho} R\left(\frac{\partial v_z}{\partial r}\right)_{r=R} + \frac{\mu}{\rho}\left(\frac{d}{dz}\left(\int_0^R \frac{\partial v_z}{\partial z} r\, dr\right) - \left(\frac{\partial v_z}{\partial z}\right)_{r=R} R\frac{dR}{dz}\right) =$

$$= -\frac{R^2}{2\rho}\cdot\frac{\partial P}{\partial z} + \frac{\mu}{\rho} R\left(\frac{\partial v_z}{\partial r}\right)_{r=R} + \frac{\mu}{\rho}\left(-\frac{d}{dz}\left(\int_0^R \frac{1}{r}\frac{\partial v_r}{\partial r} r\, dr\right) - \left(\frac{\partial v_z}{\partial z}\right)_{r=R} R\frac{dR}{dz}\right) =$$

$$= -\frac{R^2}{2\rho}\cdot\frac{\partial P}{\partial z} + \frac{\mu}{\rho} R\left(\frac{\partial v_z}{\partial r}\right)_{r=R} + \frac{\mu}{\rho}\left(-\frac{\partial}{\partial z}(R(v_r)_{r=R}) - \left(\frac{\partial v_z}{\partial z}\right)_{r=R} R\frac{dR}{dz}\right) =$$

$$= -\frac{R^2}{2\rho}\cdot\frac{\partial P}{\partial z} + \frac{\mu}{\rho} R\left(\left(\frac{\partial v_z}{\partial r}\right)_{r=R} - \left(\frac{\partial v_z}{\partial z}\right)_{r=R}\frac{dR}{dz}\right) =$$

$$= -\frac{R^2}{2\rho}\cdot\frac{\partial P}{\partial z} + \frac{\mu}{\rho} R\left(\left(\frac{\partial v_z}{\partial r}\right)_{r=R} - \left(\frac{\partial v_z}{\partial z}\right)_{r=R}\frac{dR}{dz}\right) \tag{2.7}$$

Substituting Eqs. (2.6) and (2.7) into Eq. (2.5), we obtain an equation for the z-momentum of the jet integrated over the cross-section of the channel:

$$\frac{\partial}{\partial t}\left(\int_0^R v_z r\, dr\right) + \frac{\partial}{\partial z}\int_0^R (rv_z^2\, dr) = -\frac{R^2}{2\rho}\cdot\frac{\partial P}{\partial z} + \frac{\mu}{\rho} R\left(\left(\frac{\partial v_z}{\partial r}\right)_{r=R} - \left(\frac{\partial v_z}{\partial z}\right)_{r=R}\frac{dR}{dz}\right) \rightarrow$$

$$\rightarrow \frac{\partial}{\partial t}\left(\frac{2\rho}{R^2}\int_0^R v_z r\, dr\right) + \frac{2\rho}{R^2}\frac{\partial}{\partial z}\int_0^R (rv_z^2\, dr) - \frac{2\mu}{R}\left(\left(\frac{\partial v_z}{\partial r}\right)_{r=R} - \left(\frac{\partial v_z}{\partial z}\right)_{r=R}\frac{dR}{dz}\right) = -\frac{\partial P}{\partial z} \rightarrow$$

$$\rightarrow \frac{\partial}{\partial t}\left(\frac{2\rho}{R^2}\int_0^R v_z r\, dr\right) + \frac{\partial}{\partial z}\left(\frac{2\rho}{R^2}\int_0^R (rv_z^2\, dr)\right) + \frac{4\rho}{R^3}\frac{dR}{dz}\int_0^R (rv_z^2\, dr) -$$

$$-\frac{2\mu}{R}\left(\left(\frac{\partial v_z}{\partial r}\right)_{r=R} - \left(\frac{\partial v_z}{\partial z}\right)_{r=R}\frac{dR}{dz}\right) = -\frac{\partial P}{\partial z} \rightarrow$$

$$\rightarrow \frac{2\rho}{R^2}\frac{\partial}{\partial t}\left(\int_0^R v_z r\, dr\right) + \frac{4\rho}{R^3}\frac{dR}{dz}\int_0^R (rv_z^2\, dr) - \frac{2\mu}{R}\left(\left(\frac{\partial v_z}{\partial r}\right)_{r=R} - \left(\frac{\partial v_z}{\partial z}\right)_{r=R}\frac{dR}{dz}\right) =$$

$$= -\frac{\partial}{\partial z}\left(P + \left(\frac{2\rho}{R^2}\int_0^R (rv_z^2\, dr)\right)\right) \tag{2.8}$$

Taking into account that flow, $\Phi = 2\pi\int_0^R v_z r\, dr$, is independent of z, the integration of Eq. (2.8) over z yields the following equation for the total z-momentum of the jet:

$$2\rho\left(\int_0^Z \frac{dz}{R^2}\right)\frac{\partial}{\partial t}\left(\int_0^R v_z r\, dr\right) - 2\mu\int_0^L \left\{\frac{1}{R}\left(\left(\frac{\partial v_z}{\partial r}\right)_{r=R} - \left(\frac{\partial v_z}{\partial z}\right)_{r=R}\frac{dR}{dz}\right)\right\} dz =$$

$$= -4\rho\int_0^Z \left\{\frac{1}{R^3}\frac{dR}{dz}\int_0^R (rv_z^2\, dr)\right\} dz - \left(P + \left(\frac{2\rho}{R^2}\int_0^R (rv_z^2\, dr)\right)\right)\bigg|_0^Z \tag{2.9}$$

where 0 corresponds to the entrance to the channel and $Z$ is the length of the channel that is filled with fluid, Fig. 3. Let us present the first term in the LHS of Eq. (2.9) in the following form:

$$-4\rho \int_0^Z \left\{\frac{1}{R^3}\frac{dR}{dz} \int_0^R (rv_z^2 dr)\right\} dz = -4\rho \int_0^Z \left\{\frac{1}{R^5}\frac{dR}{dz}\left\{R^2 \int_0^R (rv_z^2 dr)\right\}\right\} dz =$$

$$= \rho \int_0^Z \left\{\frac{dR^{-4}}{dz}\left(R^2 \int_0^R (rv_z^2 dr)\right)\right\} dz =$$

$$= \rho \left(\frac{1}{R^4} R^2 \int_0^R (rv_z^2 dr)\right)_0^Z + \rho \int_0^Z \left\{\frac{1}{R^4}\frac{\partial}{\partial z}\left(R^2 \int_0^R (rv_z^2 dr)\right)\right\} dz$$

Substituting this expression into Eq. (2.9), we obtain

$$\left(2\rho \int_0^Z \frac{dz}{R^2}\right)\frac{\partial}{\partial t}\left(\int_0^R v_z r dr\right) - 2\mu \int_0^Z \left\{\frac{1}{R}\left(\left(\frac{\partial v_z}{\partial r}\right)_{r=R} - \left(\frac{\partial v_z}{\partial z}\right)_{r=R}\frac{dR}{dz}\right)\right\} dz =$$

$$= -\rho \int_0^Z \left\{\frac{1}{R^4}\frac{\partial}{\partial z}\left(R^2 \int_0^R (rv_z^2 dr)\right)\right\} dz + P_0(t) - P_Z +$$

$$+ \left(\frac{\rho}{R^2}\int_0^R (rv_z^2 dr)\right)_0 - \left(\frac{\rho}{R^2}\int_0^R (rv_z^2 dr)\right)_Z \tag{2.10}$$

In Eq. (2.10), in the LHS, the first term is the inertial term and the second term is the resistant term; in the RHS, the first term is the volumetric part of dynamic pressure; $P_0(t)$ is the static pressure at the entrance to the channel; $P_Z = \frac{2\sigma\cos(\theta)}{R_Z}$ is the meniscus surface pressure, $\sigma$ is the surface tension of the liquid, $\theta$ is the solid surface - liquid contact angle, and $R_Z$ is the radius of the channel at the position of the tip, Fig. 3; and the fourth and fifth terms are the divergent parts of dynamic pressures at the entrance to the channel and at the "flat" tip of the jet respectively, Fig. 3.

In the model, we assume that the flow in the channel is fully developed,

$$v_z(r) = \frac{2\Phi}{\pi R^2}\left(1 - \left(\frac{r}{R}\right)^2\right) \tag{2.11}$$

Substituting Eq. (2.11) into Eq. (2.10), we obtain the final model equation for the total momentum of the jet in the channel:

$$\frac{\rho}{\pi}\left(\int_0^L \frac{dz}{R^2}\right)\frac{d\Phi}{dt} + \frac{8\mu\Phi}{\pi}\int_0^L \left\{\frac{1}{R^4}\left(1 + \left(\frac{dR}{dz}\right)^2\right)\right\} dz = P_0 - \frac{2\sigma\cos(\theta)}{R_L} + \frac{2\rho\Phi^2}{3\pi^2}\left(\frac{1}{R_0^4} - \frac{1}{R_L^4}\right) \tag{2.12}$$

It is worth noting that, because, in Eq. (2.10), the normalized velocity profile, $\pi R^2 v_z(r)/\Phi$, is a function of $(r/R)$ only, the volumetric part of dynamic pressure in Eq. (2.12) is zero. As we show in Section 3, in the case of a rectangular tapered channel, this is not always the case. Also, in the case of a straight channel, $dR/dz = 0$, the second term in the parentheses in the viscosity term is equal to zero. Dropping the inertial term, the first term in the RHS of Eq. (2.12), and the dynamic pressure term, the third term in LHS of Eq. (2.12), the case of $Re \ll 1$, Eq. (2.12) reduces to a standard form a steady-state flow in a circular capillary with varying cross-section [15]; in [15], $P_0 = 0$.

An equation for the length of the fluid filled portion of the channel (the length of the jet in Fig. 3) can be written as:

$$\frac{dZ}{dt} = \frac{\Phi}{\pi R_Z^2} \tag{2.13}$$

Introducing the following functions,

$$I(Z) = \frac{\rho}{\pi}\left(\int_0^Z \frac{dz}{R^2}\right), \quad K(Z) = \frac{8\mu}{\pi}\int_0^Z \left\{\frac{1}{R^4}\left(1 + \left(\frac{dR}{dz}\right)^2\right)\right\}dz, \quad \Pi_{dyn-grad}(Z) = \frac{2\rho}{3\pi^2(R(Z))^4}, \tag{2.14a}$$

$$S(Z) = \pi(R(Z))^2, \quad ST(Z) = \frac{2\sigma \cos(\theta)}{R(Z)} \tag{2.14b}$$

the set of ODEs. (2.12) and (2.13) can be written in the following form:

$$I\frac{d\Phi}{dt} + K\Phi = P_0(t) - ST + \Phi^2\left(\Pi_{dyn-grad}(0) - \Pi_{dyn-grad}(Z)\right) \tag{2.15}$$

$$\frac{dZ}{dt} = \frac{\Phi}{S} \tag{2.16}$$

where $\Phi^2\Pi_{dyn-grad}(0)$ and $\Phi^2\Pi_{dyn-grad}(Z)$ are the dynamic pressure at the entrance to the channel and at the "flat" tip of the jet, Fig. 3. Thus, the set of Eqs. (2.14) - (2.16) describes the process of filling a circular tapered channel in the fully developed flow approximation.

It has to be stressed that, in [21-27], the dynamic pressure term is in 3/2 times larger than obtained in Eq. (2.12). This happens because the dynamic pressure term in [21-27] corresponds to a plug velocity profile that is inconsistent with the fully developed flow approximation used in those models.

**3. 1D-model of flow in a tapered rectangular microchannel**

In a tapered rectangular channel, Fig. 4, a set of equations describing the flow of an incompressible fluid can be written as:

$$\frac{\partial v_z}{\partial t} + v_x \frac{\partial v_z}{\partial x} + v_y \frac{\partial v_z}{\partial y} + v_z \frac{\partial v_z}{\partial z} = -\frac{1}{\rho} \cdot \frac{\partial P}{\partial z} + \frac{\mu}{\rho}\left(\frac{\partial^2 v_z}{\partial x^2} + \frac{\partial^2 v_z}{\partial y^2} + \frac{\partial^2 v_z}{\partial z^2}\right) \tag{3.1}$$

$$\frac{\partial v_x}{\partial t} + v_x \frac{\partial v_x}{\partial x} + v_y \frac{\partial v_x}{\partial y} + v_z \frac{\partial v_x}{\partial z} = \frac{\mu}{\rho}\left(\frac{\partial^2 v_x}{\partial x^2} + \frac{\partial^2 v_x}{\partial y^2} + \frac{\partial^2 v_x}{\partial z^2}\right) \tag{3.2}$$

$$\frac{\partial v_y}{\partial t} + v_x \frac{\partial v_y}{\partial x} + v_y \frac{\partial v_y}{\partial y} + v_z \frac{\partial v_y}{\partial z} = \frac{\mu}{\rho}\left(\frac{\partial^2 v_y}{\partial x^2} + \frac{\partial^2 v_y}{\partial y^2} + \frac{\partial^2 v_y}{\partial z^2}\right) \tag{3.3}$$

$$\frac{\partial v_z}{\partial z} + \frac{\partial v_x}{\partial x} + \frac{\partial v_y}{\partial y} = 0 \tag{3.4}$$

where Eqs. (3.1), (3.2), and (3.3) are the Navier-Stokes equations correspondingly for $z$-momentum, $x$-momentum, and $y$-momentum; Eq. (3.4) is the volume conservation equation. As in Section 2, we assume the non-slip boundary conditions at the surface of the channel:

$$(v_z)_\Gamma = 0, \qquad (v_x)_\Gamma = 0, \qquad (v_y)_\Gamma = 0, \tag{3.5}$$

where $\Gamma$ is the boundary of the channel; and $Z$ as the length of the jet as shown in Fig. 4. In the model, we assume that the flow has a mirror-symmetry with respect to the $x$-axis and the $y$-axis:

$$v_z(-x, y, z) = v_z(x, y, z) \quad v_x(-x, y, z) = v_x(x, y, z) \quad v_y(-x, y, z) = -v_y(x, y, z) \tag{3.6a}$$

$$v_z(x, -y, z) = v_z(x, y, z) \quad v_x(x, -y, z) = -v_x(x, y, z) \quad v_y(x, -y, z) = v_y(x, y, z) \tag{3.6b}$$

Using the same technique as in Section 2 and taking into account that flow, $\Phi = 4 \int_0^X \int_0^Y v_z dx dy$, is independent of $z$, the integration of Eq. (3.1) over the volume of the liquid that partially fills the channel, Fig. 4, yields the following equation for the total z-momentum of the jet:

$$\left(\rho \frac{\partial}{\partial t}\left(\int_0^Y \int_0^X v_z dx dy\right)\right) \int_0^Z \frac{dz}{XY} -$$

$$- \int_0^Z \frac{\mu}{XY}\left(\int_0^Y \left(\frac{\partial v_z}{\partial x}\right)_X dy + \int_0^X \left(\frac{\partial v_z}{\partial y}\right)_Y dx - \frac{dX}{dz}\int_0^Y \left(\frac{\partial v_z}{\partial z}\right)_X dy - \frac{dY}{dz}\int_0^X \left(\frac{\partial v_z}{\partial z}\right)_Y dx\right) dz =$$

$$= P_0(t) - ST + \left(\frac{\rho}{2XY}\int_0^Y \int_0^X v_z^2 dx dy\right)_0 - \left(\frac{\rho}{2XY}\int_0^Y \int_0^X v_z^2 dx dy\right)_Z -$$

$$- \frac{\rho}{2}\int_0^Z \left(\frac{1}{X^2 Y^2}\frac{\partial}{\partial z}\left(XY \int_0^Y \int_0^X v_z^2 dx dy\right)\right) dz \tag{3.7}$$

where index "0" corresponds to the entrance to the channel and $Z$ is the length of the channel that is filled with fluid, Fig. 4; derivation of Eq. (3.7) is presented in Appendix A. In Eq. (3.7), in the LHS, the first term is the inertial term and the second term is the resistant term; in the RHS, $P_0(t)$ is the static pressure at the entrance to the channel;

$$ST = \sigma\cos(\theta)\left(\frac{1}{X} + \frac{1}{Y}\right)_Z \tag{3.8}$$

is the meniscus surface pressure, $\sigma$ is the surface tension of the liquid, $\theta$ is the solid surface - liquid contact angle, and $X_Z$ and $Y_Z$ are the channels dimensions at the position of the "flat" tip, Fig. 4; the third and fourth terms are the divergent parts of dynamic pressures at the entrance to the channel and at the "flat" tip of the jet, respectively, Fig. 4; and the fifth term is the volumetric part of the dynamic pressure.

In the case of a square channel, Eq. (3.7) reduces to the following form:

$$\left(\rho\frac{\partial}{\partial t}\left(\int_0^a\int_0^a v_z dxdy\right)\right)\int_0^Z \frac{dz}{a^2} - 2\int_0^Z \frac{\mu}{a^2}\left(\int_0^a \left(\frac{\partial v_z}{\partial y}\right)_a dx - \frac{da}{dz}\int_0^a \left(\frac{\partial v_z}{\partial z}\right)_a dx\right)dz =$$

$$= P_0(t) - ST + \left(\frac{\rho}{2a^2}\int_0^a\int_0^a v_z^2 dxdy\right)_0 - \left(\frac{\rho}{2a^2}\int_0^a\int_0^a v_z^2 dxdy\right)_Z -$$

$$-\frac{\rho}{2}\int_0^Z \left(\frac{1}{a^4}\frac{\partial}{\partial z}\left(a^2\int_0^Y\int_0^X v_z^2 dxdy\right)\right)dz \tag{3.9}$$

where $a(z)$ is the width of the channel as a function on $z$.

In the model, we assume that the flow in the channel is fully developed with $v_z$ derived from [29],

$$v_z(x,y) = \frac{\pi\Phi}{8XY}\frac{\sum_{i=0}^{i=\infty}\frac{(-1)^i}{(2i+1)^3}\cos\left(\frac{(2i+1)\pi x}{2X}\right)\left(1-\frac{\cosh\left(\frac{(2i+1)\pi y}{2X}\right)}{\cosh\left(\frac{(2i+1)\pi Y}{2X}\right)}\right)}{\sum_{i=0}^{i=\infty}\frac{1}{(2i+1)^4}\left(1-\frac{2X}{(2i+1)\pi Y}tgh\left(\frac{(2i+1)\pi Y}{2X}\right)\right)} \tag{3.10}$$

where $\Phi$ is the flow in the channel,

$$\frac{\Phi}{4} = \int_0^Y\int_0^X v_z dxdy \tag{3.11}$$

Substituting Eq. (3.10) into Eq. (3.7), we obtain the final model equation for the total momentum of the jet in the channel, which can be presented in the following form (Appendix B):

$$\frac{\rho}{4}\frac{d\Phi}{dt}\int_0^Z \frac{dz}{XY} + \frac{\mu\pi^2\Phi}{16}\int_0^Z \left(F_{unity}(\xi) + \left(\frac{dX}{dz}\right)^2 F_{xx}(\xi) + \left(\frac{dY}{dz}\right)^2 F_{yy}(\xi)\right)\frac{dz}{Y^2X^2} =$$

$$= P_0(t) - ST + \frac{\pi^2\rho\Phi^2}{256}\left(\frac{U(\xi)}{X^2Y^2}\right)_0 - \frac{\pi^2\rho\Phi^2}{256}\left(\frac{U(\xi)}{X^2Y^2}\right)_Z - \frac{\pi^2\rho\Phi^2}{256}\int_0^Z \frac{dU(\xi)}{d\xi}\frac{d\xi}{dz}\frac{dz}{X^2Y^2} \tag{3.12}$$

where

$$U(\xi) = \frac{\sum_{i=0}^{i=\infty} \frac{1}{(2i+1)^6} \left( \int_0^1 \left( 1 - \frac{\cosh\left(\frac{(2i+1)\pi\xi}{2}\hat{y}\right)}{\cosh\left(\frac{(2i+1)\pi\xi}{2}\right)} \right)^2 d\hat{y} \right)}{\left( \sum_{i=0}^{i=\infty} \frac{1}{(2i+1)^4} \left( 1 - \frac{2}{(2i+1)\pi\xi} tgh\left(\frac{(2i+1)\pi\xi}{2}\right) \right) \right)^2} \tag{3.13}$$

$$F_{unity}(\xi) = \frac{\sum_{i=0}^{i=\infty} \frac{\xi}{(2i+1)^2}}{\sum_{i=0}^{i=\infty} \frac{1}{(2i+1)^4} \left( 1 - \frac{2}{(2i+1)\pi\xi} tgh\left(\frac{(2i+1)\pi\xi}{2}\right) \right)} \tag{3.14}$$

$$F_{xx}(\xi) = \frac{\sum_{i=0}^{i=\infty} \frac{\xi}{(2i+1)^2} - \sum_{i=0}^{i=\infty} \frac{2}{(2i+1)^3\pi} tgh\left(\frac{(2i+1)\pi\xi}{2}\right)}{\sum_{i=0}^{i=\infty} \frac{1}{(2i+1)^4} \left( 1 - \frac{2}{(2i+1)\pi\xi} tgh\left(\frac{(2i+1)\pi\xi}{2}\right) \right)} \tag{3.15}$$

$$F_{yy}(\xi) = \frac{\sum_{i=0}^{i=\infty} \frac{2}{(2i+1)^3\pi} tgh\left(\frac{(2i+1)\pi\xi}{2}\right)}{\sum_{i=0}^{i=\infty} \frac{1}{(2i+1)^4} \left( 1 - \frac{2}{(2i+1)\pi\xi} tgh\left(\frac{(2i+1)\pi\xi}{2}\right) \right)} \tag{3.16}$$

$$\xi = \frac{Y}{X} \tag{3.17}$$

Functions $K(\xi)$, $F_{unity}(\xi)$, $F_{dXdz}(\xi)$ have the following properties:

$$U\left(\frac{1}{\xi}\right) = U(\xi), \quad F_{unity}\left(\frac{1}{\xi}\right) \equiv F_{unity}(\xi), \quad \text{and} \quad F_{yy}(\xi) = F_{xx}\left(\frac{1}{\xi}\right) \tag{3.18}$$

It should be noted that, in case where the ratio of $Y$ to $X$, Fig. 4, is independent of z, $d\xi/dz = 0$, the volumetric part of the dynamic pressure is zero, see the last term in the RHS of Eq. (3.12). Also, in the case of a straight channel, $dX/dz = dY/dz = 0$, and the second and the third terms in the parentheses in the viscosity term are equal to zero. In the case of $Re \ll 1$, the inertial and dynamic pressures terms can be dropped in Eq. (3.12) that reduces this equation to a form of a steady-state flow in a rectangular capillary channel with varying cross-section.

An equation for the length of the fluid filled portion of the channel (the length of the jet in Fig. 4) can be written as

$$\frac{dZ}{dt} = \frac{\Phi}{4X_Z Y_Z} \tag{3.19}$$

Introducing the following functions,

$$I = \frac{\rho}{4}\int_0^Z \frac{dz}{XY}, \quad K = \frac{\mu\pi^2}{16}\int_0^Z \left(F_{unity}(\xi) + \left(\frac{dX}{dz}\right)^2 F_{dXdz}(\xi) + \left(\frac{dY}{dZ}\right)^2 F_{dYdz}(\xi)\right)\frac{dz}{Y^2 X^2}, \quad (3.20a)$$

$$\Pi_{dyn-grad} = \frac{\pi^2 \rho U(\xi)}{256 X^2 Y^2}, \quad \Pi_{dyn-vol} = \frac{\pi^2 \rho}{256}\int_0^Z \frac{dU(\xi)}{d\xi}\frac{d\xi}{dz}\frac{dz}{X^2 Y^2} \quad (3.20b)$$

$$S = 4X_Z Y_Z, \quad ST = \sigma \cos(\theta)\left(\frac{1}{X_Z} + \frac{1}{X_Z}\right) \quad (3.20c)$$

the set of ODEs. (3.12) and (3.19) can be written in the following form:

$$I\frac{d\Phi}{dt} + K\Phi = P_0(t) + \Pi_{dyn-grad}(0)\Phi^2 - ST + \Phi^2 \Pi_{dyn-grad}(Z) - \Pi_{dyn-vol}\Phi^2 \quad (3.21)$$

$$\frac{dZ}{dt} = \frac{\Phi}{S} \quad (3.22)$$

where $\Pi_{dyn-grad}(0)\Phi^2$ and $\Pi_{dyn-grad}(Z)\Phi^2$ are the gradient parts of dynamic pressure at the entrance to the channel and at the "flat" tip of the jet, Fig. 4. Thus, a set of ODEs. (3.21) and (3.22) describes the process of the filling of a rectangular tapered channel in the fully developed flow approximation.

## 4. Modeling of three-channel Y-shape capillary network, circular channels

Let us first consider a 1D-full model for a Y-shape capillary network with circular channels, Fig. 2. In this model, we take into account the inertance and dynamic pressure terms. This model can be divided into two parts: Case 1. The fluid fills channel 1 only. Case 2. The fluid reaches the $Y$ – junction and starts filling channels 2 and 3. In Subsection 3, we construct a creeping flow Y-shape network model. In this model we assume that $Re \ll 1$ and, therefore, inertance and dynamic pressure terms are negligibly small and are dropped in Eq. (2.12); we call this model a reduced model.

### *1. Full model, case of filling of channel 1*

In the full model, in Case 1, a set of the ordinary differential equations describing the filling of channel 1 only consists of Eqs. (2.15) and (2.16):

$$I_1 \frac{d\Phi_1}{dt} + K_1 \Phi_1 = P_0(t) - ST_1 - \Pi_{dyn-grad,1}(Z)\Phi_1^2 \quad (4.1)$$

$$\frac{dZ_1}{dt} = \frac{\Phi}{S_1} \quad (4.2)$$

where functions $I_1$, $K_1$, $ST_1$, $S_1$, and $\Pi_{dyn-grad,1}$ are presented by Eqs. (2.14a) and (2.14b); and index "1" corresponds to the first channel in Fig. 2. In Eq. (4.1), we have taken into account that the gradient part of dynamic pressure at the interface between the reservoir and channel 1, Fig. 2, is small and, therefore, set to zero. A set of initial boundary conditions for ODEs. (4.1) and (4.2) is:

$$Z_1(t=0) = 0 \quad \text{and} \quad \Phi_1(t=0) = \sqrt{\frac{3\pi^2(R(0))^4}{2\rho}\left(P_0(t=0) - \frac{2\sigma\cos(\theta)}{R(0)}\right)} \tag{4.3}$$

The initial boundary conditions for $\Phi_1$, Eq. (4.3), has been obtained from Eq. (4.1) by taking into account that the LHS of Eq. (4.1) at $t=0$ is equal to zero. ODEs. (4.1) and (4.2) have to be integrated until $Z_1$ reaches $L_1$ where $L_1$ is the length of channel 1, Fig. 2.

*2. Full model, case of filling of channels 2 and 3*

In the case of three channels, 1, 2, and 3, a set of equations describing the filling of channels 2 and 3 (channel 1 is completely filled, Fig. 2) can be written as

$$I_1 \frac{d\Phi_1}{dt} + K_1 \Phi_1 = P_0 - P_1 - \Pi_{dyn-grad,1}\Phi_1^2 \tag{4.4}$$

$$I_2 \frac{d\Phi_2}{dt} + K_2 \Phi_2 = P_1 + \Pi_{dyn-grad,1}\Phi_1^2 - \{ST\}_2 - \{\Pi_{dyn-grad}\Phi^2\}_2 \tag{4.5}$$

$$I_3 \frac{d\Phi_3}{dt} + K_3 \Phi_3 = P_1 + \Pi_{dyn-grad,1}\Phi_1^2 - \{ST\}_3 - \{\Pi_{dyn-grad}\Phi^2\}_3 \tag{4.6}$$

$$\Phi_1 = \Phi_2 + \Phi_3 \tag{4.7}$$

$$\frac{dZ_2}{dt} = \frac{\Phi_2}{S_2} \tag{4.8}$$

$$\frac{dZ_3}{dt} = \frac{\Phi_3}{S_3} \tag{4.9}$$

where

$$\{ST\}_2 = ST_1(L_1) + \frac{ST_2(Z_2) - ST_1(L_1)}{R_2(0)} \cdot \begin{cases} Z_2 & \text{for } Z_2 \leq R_2(0) \\ R_2(0) & \text{for } Z_2 > R_2(0) \end{cases} \tag{4.10}$$

$$\{ST\}_3 = ST_1(L_1) + \frac{ST_3(Z_3) - ST_1(L_1)}{R_3(0)} \cdot \begin{cases} Z_3 & \text{for } Z_3 \leq R_3(0) \\ R_3(0) & \text{for } Z_3 > R_3(0) \end{cases} \tag{4.11}$$

$$\{\Pi_{dyn-grad}\Phi^2\}_2 = \Pi_{dyn-grad,1}\Phi_1^2 + \frac{\Pi_{dyn-grad,2}\Phi_2^2 - \Pi_{dyn-grad,1}\Phi_1^2}{R_2(0)} \cdot \begin{cases} Z_2 & \text{for } Z_2 \leq R_2(0) \\ R_2(0) & \text{for } Z_2 > R_2(0) \end{cases} \tag{4.12}$$

$$\{\Pi_{dyn-grad}\Phi^2\}_3 = \Pi_{dyn-grad,1}\Phi_1^2 + \frac{\Pi_{dyn-grad,3}\Phi_3^2 - \Pi_{dyn-grad,1}\Phi_1^2}{l_3} \cdot \begin{cases} Z_3 & for\ Z_3 \leq R_3(0) \\ R_3(0) & for\ Z_3 > R_3(0) \end{cases} \quad (4.13)$$

$R_2(0)$ is the entrance radius of section 2, and $R_3(0)$ is the entrance radius of Section 3, Fig. 2. Initial boundary conditions for ODEs (4.5), (4.6), (4.8) and (4.9) are:

$$Z_2(t_{1,end}) = 0 \quad and \quad \Phi_2(t_{1,end}) = \Phi_1(t_{1,end})\frac{S_2(0)}{S_3(0)+S_2(0)} \quad (4.14)$$

$$Z_3(t_{1,end}) = 0 \quad and \quad \Phi_3(t_{1,end}) = \Phi_1(0)\frac{S_3(0)}{S_3(0)+S_2(0)} \quad (4.15)$$

Here $P_1$ and $\Pi_{dyn-grad,1}\Phi_1^2$ are the static and dynamic fluid pressures, respectively, at the interface between channels 1, 2, and 3 (see Fig. 2); constants $I_1 = I_1(L_1)$, $K_1 = K_1(L_1)$, $\Pi_{dyn-grad,1} = \Pi_{dyn-grad,1}(L_1)$; functions $I_2$, $K_2$, $\Pi_{dyn-grad,2}$, $S_2$, $ST_2$, $I_3$, $K_3$, $\Pi_{dyn-grad,3}$, $S_3$ $ST_3$ are presented by Eqs. (2.14a) and (2.14b) with indexes "2" and "3" for channels 2 and 3, respectively; and $t_{1,end}$ is the time when the liquid starts filling of channels 2 and 3. ODEs. (4.4) - (4.9) have to be integrated until $Z_2$ reaches $L_2$ or $Z_3$ reaches $L_3$, where $L_2$ and $L_3$ are, correspondingly, the lengths of channels 2 and 3, Fig. 2. It is worth reminding that, in the case of circular channels, the terms responsible for volumetric part of dynamic pressure are equal to zero and, therefore, absent in Eqs. (4.4) – (4.6), see Section 2.

Eqs. (4.4) – (4.6) describe the momentum conservation law, correspondingly, in channels 1, 2, and 3; Eq. (4.7) states that the volume is conserved in incompressible liquids; and Eqs. (4.8) and (4.9) describe the movement of the "flat" meniscuses in channels 2 and 3, respectively.

To describe the transition process of filling of the Y-shape junction, Fig. 2, in Eqs. (4.5) and (4.6), we have introduced the smoothing surface tension pressures $\{ST\}_2$ and $\{ST\}_3$, Eq. (4.10) - (4.11), and smoothing gradient part of dynamic pressures $\{\Pi_{dyn-grad}\Phi^2\}_2$ and $\{\Pi_{dyn-grad}\Phi^2\}_2$, Eqs. (4.12) – (4.13), with the characteristic smoothing lengths of $R_2(0)$ and $R_3(0)$. Indeed, at $t = t_{1,end}$, the LHSs of Eqs. (4.5) and (4.6) are equal to zero. Therefore, using $ST_2$ and $\Pi_{dyn-grad,2}\Phi_2^2$ in Eq. (4.5), and $ST_3$ and $\Pi_{dyn-grad,3}\Phi_3^2$ in Eqs. (4.6) instead of $\{ST\}_2$, $\{ST\}_3$, $\{\Pi_{dyn-grad}\Phi^2\}_2$ and $\{\Pi_{dyn-grad}\Phi^2\}_3$ would lead to uncertainty in the value of $P_1(t_{1,end}) + \Pi_{dyn-grad,1}\Phi_1^2(t_{1,end})$.

In the case where the inlet cross-sections of channels 2 and 3 are different, the meniscus position at the channel with the larger inlet cross-section can be arrested for some period of time [18]. In our simulations, we observed this phenomenon as a negative flow from the channel with the larger inlet cross-section, a result that makes no sense. Therefore, in the code, we also use the following conditions:

$$\text{if } \Phi_2 < 0 \text{ then } \Phi_2 = 0 \quad \text{and} \quad \text{if } \Phi_3 < 0 \text{ then } \Phi_3 = 0 \tag{4.16}$$

As we show in Section 8, these conditions indeed describe the meniscus arrest.

It is worth noting that, in our code, we use a slightly modified set of equations in which we have eliminated $P_1(t)$ and $\Phi_1$ from Eqs. (4.4) – (4.9) to get a set of four ordinary differential equations for $\Phi_2$, $\Phi_3$, $Z_2$, and $Z_3$.

*3. Reduced model*

In Case 1, for creeping flow (the reduced model), $Re \ll 1$, the inertance and dynamic pressure terms are negligibly small and can be dropped from Eq. (4.1); that yields to the following set of creping flow equations:

$$K_1 \Phi = P_0(t) - ST_1 \qquad \frac{dZ_1}{dt} = \frac{\Phi}{S_1} \tag{4.17}$$

At small times, an analytical solution of these equations is:

$$Z_1 = \frac{R_1}{2}\left(\frac{(P_0 - ST_1)t}{\mu\left(1 + \frac{dR_1}{dz}^2\right)}\right)^{1/2} \qquad \Phi_1 = \pi R_1^2 \frac{Z_1}{2t} \tag{4.18}$$

where $R_1$ and $\frac{dR_1}{dz}$ are taken at the entrance to the channel and $P_0$ is the pressure at $t = 0$. As one can see from Eqs. (4.18), as $t \to 0$, $\Phi_1 \to \infty$. To avoid this, in the code, we use the time integration step $\Delta t$ as $t$ in Eqs. (4.18) to obtain the initial values of $Z_1$ and $\Phi_1$. Varying the integration time step, we observed a conversion of solutions for times which are much larger than time steps.

In Case 2, of filling of channels 2 and 3 (channel 1 is completely filled), a set of the creeping flow equations can be obtained by dropping the inertance and the dynamic pressure terms in Eqs. (4.5) and (4.6):

$$K_1 \Phi_1 = P_0 - P_1 \tag{4.19}$$

$$K_2 \Phi_2 = P_1 - \{ST\}_2 \tag{4.20}$$

$$K_3 \Phi_3 = P_1 - \{ST\}_3 \tag{4.21}$$

$$\Phi_1 = \Phi_2 + \Phi_3 \tag{4.22}$$

$$\frac{dZ_2}{dt} = \frac{\Phi_2}{S_2} \tag{4.23}$$

$$\frac{dZ_3}{dt} = \frac{\Phi_3}{S_3} \tag{4.24}$$

where $\{ST\}_2$ and $\{ST\}_3$ are given correspondingly by Eqs. (4.10) and (4.11). The initial boundary conditions for this set of equations are the same as in the case of the full model, Eqs. (4.14) and (4.15); and, in the code, Eq. (4.16) has to be used as well.

## 5. Modeling of three-channel Y-shape capillary network, rectangular channels

As Section 4, this section is also divided into three parts. In the first two subsections we construct a full model describing the filling of a Y-shape capillary network consisting with three rectangular channels, Fig. 2. In this model, we take into account the inertance and the dynamic pressure. Then, in the third subsection, we construct a creeping flow three-channel Y-shape network model; we call this model a reduced model.

### *1. Full model, case of filling of channel 1 only*

In Case 1, a set of the ordinary differential equations describing the filling of Section 1 only consists of Eqs. (3.21) and (3.22):

$$I_1 \frac{d\Phi}{dt} + K_1 \Phi = P_0(t) - ST_1 - \Pi_{dyn-grad,1} \Phi^2 - \Pi_{dyn-vol,1} \Phi^2 \tag{5.1}$$

$$\frac{dZ_1}{dt} = \frac{\Phi}{S_1} \tag{5.2}$$

where functions $I_1$, $K_1$, $ST_1$, $S_1$, $\Pi_{grad-dyn,1}$, and $\Pi_{vol-dyn,1}$ are presented by Eqs. (3.20); and index "1" corresponds to the first channel. As in Section 4, in Eq. (5.1), we dropped the dynamic pressure term at the entrance to channel 1. A set of initial boundary conditions for ODEs. (5.1) and (5.2) is:

$$Z_1(t=0) = 0 \quad \text{and} \quad \Phi_1(t=0) = \left( \frac{P_0(t=0) - \sigma \cos(\theta)\left(\frac{1}{X(0)} + \frac{1}{Y(0)}\right)}{\Pi_{dyn-grad,1}(0)} \right)^{1/2} \tag{5.3}$$

The initial boundary condition for $\Phi_1$, Eq. (5.3), was obtained from Eq. (5.1) by taking into account that the LHS of Eq. (5.1) and $\Pi_{dyn-vol,1}$ at $t = 0$ are equal to zero. ODEs. (5.1) and (5.2) have to be integrated until $Z_1$ reaches $L_1$ where $L_1$ is the length of channel 1, Fig. 2.

*Case 2. Full model, case of filling of channels 2 and 3*

In Case 2, a set of equations describing the filling of channels 2 and 3 (channel 1 is completely filled, Fig. 2) can be written as

$$I_1 \frac{d\Phi_1}{dt} + K_1 \Phi_1 = P_0 - P_1 - \Pi_{dyn-grad,1}\Phi_1^2 - \Pi_{dyn-vol,1}\Phi_1^2 \tag{5.4}$$

$$I_2 \frac{d\Phi_2}{dt} + K_2 \Phi_2 = P_1 + \Pi_{dyn-grad,1}\Phi_1^2 - \{ST\}_2 - \{\Pi_{dyn-grad}\Phi^2\}_2 - \Pi_{dyn-vol,2}\Phi_2^2 \tag{5.5}$$

$$I_3 \frac{d\Phi_3}{dt} + K_3 \Phi_3 = P_1 + \Pi_{dyn-grad,1}\Phi_1^2 - \{ST\}_3 - \{\Pi_{dyn-grad}\Phi^2\}_3 - \Pi_{dyn-vol,3}\Phi_3^2 \tag{5.6}$$

$$\Phi_1 = \Phi_2 + \Phi_3 \tag{5.7}$$

$$\frac{dZ_2}{dt} = \frac{\Phi_2}{S_2} \tag{5.8}$$

$$\frac{dZ_3}{dt} = \frac{\Phi_3}{S_3} \tag{5.9}$$

where

$$\{ST\}_2 = ST_1(L_1) + \frac{ST_2(Z_2) - ST_1(L_1)}{l_2} \cdot \begin{cases} Z_2 & \text{for } Z_2 \leq l_2 \\ l_2 & \text{for } Z_2 > l_2 \end{cases} \tag{5.10}$$

$$\{ST\}_2 = ST_1(L_1) + \frac{ST_3(Z_3) - ST_1(L_1)}{l_3} \cdot \begin{cases} Z_3 & \text{for } Z_3 \leq l_3 \\ l_3 & \text{for } Z_3 > l_3 \end{cases} \tag{5.11}$$

$$\{\Pi_{dyn-grad}\Phi^2\}_2 = \Pi_{dyn-grad,1}\Phi_1^2 + \frac{\Pi_{dyn-grad,2}\Phi_2^2 - \Pi_{dyn-grad,1}\Phi_1^2}{l_2} \cdot \begin{cases} Z_2 & \text{for } Z_2 \leq l_2 \\ l_2 & \text{for } Z_2 > l_2 \end{cases} \tag{5.12}$$

$$\{\Pi_{dyn-grad}\Phi^2\}_3 = \Pi_{dyn-grad,1}\Phi_1^2 + \frac{\Pi_{dyn-grad,3}\Phi_3^2 - \Pi_{dyn-grad,1}\Phi_1^2}{l_3} \cdot \begin{cases} Z_3 & for\ Z_3 \leq l_3 \\ l_3 & for\ Z_3 > l_3 \end{cases} \quad (5.13)$$

$$l_2 = min(X_2(z=0), Y_2(z=0)) \quad and \quad l_3 = min(X_3(z=0), Y_3(z=0)) \quad (5.14)$$

Initial boundary conditions for ODEs (5.5), (5.6), (5.8) and (5.9) are:

$$Z_2(t_{1,end}) = 0 \quad and \quad \Phi_2(t_{1,end}) = \Phi_1(t_{1,end})\frac{S_2(0)}{S_3(0)+S_2(0)} \quad (5.15)$$

$$Z_3(t_{1,end}) = 0 \quad and \quad \Phi_3(t_{1,end}) = \Phi_1(0)\frac{S_3(0)}{S_3(0)+S_2(0)} \quad (5.16)$$

Here $P_1$ and $\Pi_{dyn-grad,1}\Phi_1^2$ are the static and gradient part of dynamic pressures, respectively, at the interface between channels 1, 2, and 3 (see Fig. 2); constants $I_1 = I_1(L_1)$, $K_1 = K_1(L_1)$, $\Pi_{dyn-grad,1} = \Pi_{dyn-grad,1}(L_1)$ and $\Pi_{dyn-vol,1} = \Pi_{dyn-vol,1}(L_1)$; functions $I_2$, $K_2$, $\Pi_{dyn-grad,2}$, $\Pi_{dyn-vol,2}$, $S_2$, $ST_2$, $I_3$, $K_3$, $\Pi_{dyn-grad,3}$, $\Pi_{dyn-vol,3}$, $S_3$, $ST_3$ are presented by Eqs. (3.20) with indexes "2" and "3" for channels 2 and 3, respectively; and $t_{1,end}$ is the time when the liquid starts filling channels 2 and 3. ODEs. (5.4) - (5.9) have to be integrated until $Z_2$ reaches $L_2$ or $Z_3$ reaches $L_3$, where $L_2$ and $L_3$ are, correspondingly, the lengths of channels 2 and 3, Fig. 2.

Eqs. (5.4) – (5.6) describe the momentum conservation law, correspondingly, in channels 1, 2, and 3; Eq. (5.7) states that that the volume is conserved in incompressible liquids; and Eqs. (5.8) and (5.9) describe the movement of the "flat" meniscuses in channels 2 and 3, respectively.

As in Section 4, to describe the transition process of filling of the Y-shape junction, Fig. 2, in Eqs. (5.5) and (5.6), we have introduced the smoothing surface tension pressures $\{ST\}_2$ and $\{ST\}_3$, Eq. (5.10) - (5.11), and smoothing of the gradient parts of dynamic pressures $\{\Pi_{dyn-grad}\Phi^2\}_2$ and $\{\Pi_{dyn-grad}\Phi^2\}_3$, Eqs. (5.12) – (5.13), with the characteristic smoothing lengths of $l_2$ and $l_3$ given by Eq. (5.14) that have perfect physical sense.

To avoid the non-physical solutions produced by the model in the case of the flow arrest [18], in the code, we also use the following conditions:

$$if\ \Phi_2 < 0\ then\ \Phi_2 = 0 \quad and \quad if\ \Phi_3 < 0\ then\ \Phi_3 = 0 \quad (5.17)$$

which is the same as given by Eq. (4.16).

It is worth noting that, in our code, we use a slightly modified set of equations in which we have eliminated $P_1(t)$ and $\Phi_1$ from Eqs. (5.4) – (5.9) to get a set of four ordinary differential equations for $\Phi_2$, $\Phi_3$, $Z_2$, and $Z_3$.

*3. Reduced model*

Now let us consider a creeping flow Y-shape network model, the reduced model. In this model, in Case 1, a set of equations describing the filling of channel 1, Fig. 2, is:

$$K_1\Phi = P_0(t) - ST_1 \qquad \frac{dZ_1}{dt} = \frac{\Phi}{S_1} \qquad (5.18)$$

Eq. (5.18) for $\Phi$ has been obtained by dropping the inertance and dynamic pressure terms in Eq. (5.1). At small times, this set of equations reduces to this set of equations:

$$\frac{\mu\pi^2}{16}\left(\frac{F_{unity}(\xi)+\left(\frac{dX}{dz}\right)^2 F_{dXdz}(\xi)+\left(\frac{dY}{dz}\right)^2 F_{dYdz}(\xi)}{Y^2 X^2}\right)_{z=0} z_1\Phi_1 = P_0(t) - ST_1(z=0) \qquad \frac{dZ_1}{dt} = \frac{\Phi_1}{S_1(z=0)} \qquad (5.19)$$

where $z=0$ corresponds to the entrance of channel 1, Fig. 2. An analytical solution of these equations is:

$$Z_1 = \frac{\sqrt{32(P_0(t)-ST_1(z=0))t}}{\sqrt{\mu\pi^2\left(\frac{F_{unity}(\xi)+\left(\frac{dX}{dz}\right)^2 F_{dXdz}(\xi)+\left(\frac{dY}{dz}\right)^2 F_{dYdz}(\xi)}{Y^2 X^2}S_1\right)_{z=0}}} \qquad \Phi_1 = S_1(z=0)\frac{Z_1}{2t} \qquad (5.20)$$

As one can see from Eq. (5.20), as $t \to 0$, $\Phi_1 \to \infty$. To avoid this, in the code, we use the time integration step $\Delta t$ as $t$ in Eqs. (5.20) to obtain the initial values of $Z_1$ and $\Phi_1$. Varying the integration time step, we observed a conversion of solutions for times which are much larger than time steps. This was expected, because for $t \sim \frac{\rho S}{\mu} \gg \Delta t$ (see Eq. (3.12)), the dynamic of the filling process becomes weakly dependent on chosen initial boundary conditions.

In Case 2, a set of the creeping flow equations can be obtained from the set of Eqs. (5.4) – (5.9) by dropping the inertance and the dynamic pressure terms in Eqs. (5.4) – (5.6):

$$K_1\Phi_1 = P_0 - P_1 \qquad (5.21)$$

$$K_2\Phi_2 = P_1 - \{ST\}_2 \qquad (5.22)$$

$$K_3 \Phi_3 = P_1 - \{ST\}_3 \tag{5.23}$$

$$\Phi_1 = \Phi_2 + \Phi_3 \tag{5.24}$$

$$\frac{dZ_2}{dt} = \frac{\Phi_2}{S_2} \tag{5.25}$$

$$\frac{dZ_3}{dt} = \frac{\Phi_3}{S_3} \tag{5.26}$$

where $\{ST\}_2$, are $\{ST\}_3$ are given by Eqs. (5.10), (5.11), and (5.14); the initial boundary conditions for this set of equations are the same as in the case of the full model, Eqs. (5.15) and (5.16); and Eq. (5.17) has to be used to avoid non-physical solutions if the flow arrest at Y-shape junctions appears [18].

## 6. Transition length from the plugged flow to the fully developed flow

In [28], the authors construct a model describing the transition of the plug flow to the fully developed flow in the case of a stationary liquid flow in a circular uniform pipe connected to a reservoir. In that model, the authors assume that the pipe is much longer than the transition length and completely filled by the liquid. In our model, we assume that flow is fully developed if the radius of the plugged area in model [28] is less than 0.1 of the radius of the pipe. Then, using Table 1 in [28], we can present the transition length [28] in the following form:

$$l_{plug \to developed} = \frac{0.1056 \rho \Phi}{\pi \mu} \tag{6.1}$$

Unlike in [28], in our situation, the flow and the length of the jet in the channel change with time and the channel is tapered, Fig. 3. Therefore, in the model, we determine the transition length from the plug-flow to the fully-developed-flow, $l_{trans}$, as the length of the jet when it becomes equal to the transition length calculated by Eq. (6.1). When the length of the jet is larger than $l_{trans}$, the flow is assumed to be fully developed. Thus, if $l_{trans}$ is much smaller than the length of the channel, then our assumption of fully developed flow is valid most of the time during the filling of this channel, and, therefore, the predicted time of filling of this channel is quite reasonable.

In rectangular models, we assume that the flow is fully developed if the diameter of the plugged area in model [28] is less than 0.1 of the hydraulic diameter of the cross-section of the channel,

$$D_{hydr} = \frac{4S}{P_c} \tag{6.2}$$

where $S = 4XY$ is the area and $P_c = 4X + 4Y$ is the perimeter of the channel cross-section, Fig. 4. Then, using Table 1 in [28], we can present the transition length [28] in the following form:

$$l_{plug \to developed} = \frac{0.1056 \rho \Phi}{\mu} \frac{XY}{(X+Y)^2} \tag{6.3}$$

As in the case of circular channels, we determine the transition length from the plug-flow to the fully-developed-flow, $l_{trans}$, as the length of the jet when it becomes equal to the transition length calculated by Eq. (6.3). When the length of the jet is larger than $l_{trans}$, the flow is assumed to be fully developed.

## 7. Numerical Results

To illustrate our 1D-model of flow in tapered microchannels, we simulated the flow in the two Y-shape junctions, Fig. 2, in the case of rectangular channels; the geometries of the junction are shown in Figs. 5. In the model, we use the following properties of the liquid (water): $\rho = 1000$ kg/m³, $\sigma = 0.072$ N/m, $\mu = 0.00089$ kg/(s·m). The contact angle $\theta$ was chosen as 180 degree, corresponding to a perfect wetting surface. To compare the creeping flow Y-shape junction model with the full model, we performed two tests: $P_0(t) = 0$, and $P_0(t) = 10^5$ Pa. The results of the simulations are presented in Figs. 6 -11. In the code, we used the sixth order polynomial approximation functions, Appendix C, for $U(\xi)$, $F_{unity}(\xi)$, $F_{xx}(\xi)$, $F_{yy}(\xi)$, Eqs. (3.13) – (3.16), and $dU(\xi)/d\xi$ obtained with Mathematica in the range of [0.2 – 5] which is larger than the variation of $\xi$ in the junction models shown in Fig. 5.

Figs. 6 and 7, the full model, correspond to a pure capillary flow regime where $P_0$, the external pressure at the entrance to channel 1, Fig. 2, is set to zero. Fig. 6 shows the filling process of the junction and, in Fig. 7, we magnify the area where liquid begins flowing in channels 2 and 3. As one can see from Fig. 7, the flow in channel 3 is arrested for 22 microseconds and then resumes. In Fig. 7, the sudden curve flattening of the meniscus position for channel 3 is due to a sharp decrease in the flow from its initial value given by Eq. (5.16) to zero because of Eq. (5.17). Since, in our simulation, the characteristic time of flow adjustment to initial boundary conditions, $t_{3,IBC-adj}$ in Fig. 7, is much smaller compared to the characteristic times of filling of the channels, the obtained results are not very sensitive to initial boundary conditions.

As one can see from Figs. 6 and 8, in the capillary flow regime, both models produce almost identical results. This indicates that, in this regime, the Reynolds numbers are much smaller than unity and flows are creeping. In the reduced model, $t_{3,arrest}$ is 57 microseconds, Fig. 9. It should be stressed that since, in the reduced model, the inertance term is dropped, the flow in channel 3 changes instantaneously from its initial boundary condition value to its plateau value; in the simulation $t_{3,IBS-adj}$ is one time step. That is why the plateau value of the meniscus position in Fig. 9 is much smaller than in Fig. 7.

In Fig. 10, we represent the results of the flow simulation, the full model with $P_0(t) = 0$, where the entrance cross-section area of channel 2 is much smaller than the entrance cross-section area of channel 3, geometry (b) in Fig. 5. Because of the small entrance area of channel 2, the rate of filling of channel 2 in Fig. 10a, is smaller than the rate of filling of this channel in Fig. 6. As one can see from Fig. 10a, the arrest flow in channel 3 is absent although the surface tension force at the entrance of channel 2 is 4.64 times larger than channel 3. A reason for this is because the flow resistance in channel 2 is much larger than in channel 3 so that the initial flow in channel 3 cannot be stopped by the differences in the surface tension pressures in channels 2 and 3, Fig. 10b. Thus, this simulation demonstrates that, in the capillary flow regime, the differences in surface tension pressures at asymmetric Y-shape junctions does not guarantee an appearance of the flow arrest in the channel with a larger cross-section [18].

With an increase in the external pressure $P_0$, the models start deviating from each other. As one can see in Fig. 11, geometry (a) in Fig. 5, at $P_0 = 10^5$ Pa, the differences between the models become large. Since, in the reduced model, the inertia and dynamic pressure terms are dropped, in this model, the process of filling of channel 1 is faster than in the full model. In the full model, Fig. 10, channel 1 fills in 138 microseconds, and, in the reduced model, in about 104 microseconds. This indicates that the Reynold numbers during the filling of channel 1 are larger than unity.

However, with further flowing liquid in channels 2 and 3, the flow in the channels decreases and, consequently, the Reynolds numbers in the channels decrease, becoming less than unity and, therefore, the differences in the rates of filling of the these channels calculated by both models become less distinct, Fig. 11.

With an increase in the external pressure $P_0$, the surface tension becomes less pronounced and, therefore, the meniscus arrest time decreases; at some point, at large $P_0$, the meniscus arrest disappears, Fig. 11.

In the full-model simulations with geometry (a), Fig. 5, the transition lengths, $l_{plug \rightarrow developed}$, from plug-flow to fully-developed-flow are 215 and 771 μm, correspondingly, for $P_0 = 0$ and $P_0 = 10^5$ Pa. Since the length of channel 1 is 1500 μm, the model describes reasonably well the process of filling channel 1 in the pure capillary regime, but, apparently, not so well for $P_0 = 10^5$ Pa. However, since the distance that the liquid has to go to fill channels 2 and 3 is much larger than $l_{plug \rightarrow developed}$ and, more importantly, the time needed for the flow to reach the ends of channels 2 or 3 is much larger than the time for the jet to cover the $l_{plug \rightarrow developed}$ distance, Figs. 6 and 11, the numerical results presented in this section are quite realistic.

The simulations of three-channel Y-shape junctions with circular channels are similar to the case of the rectangular channels presented above and, therefore, not shown in this article.

## 8. Conclusion

In this article, we constructed two 1D-models of laminar liquid flows in circular and rectangular tapered microchannel, the full model and the reduced model. The full model includes the inertance and dynamic pressure terms and, in the reduced model (the creeping flow cartridge model), these terms are dropped. In both model, the flows are assumed to be fully developed.

To demonstrate the method we applied the rectangular models to three-channel Y-shape junctions. We showed that in the pure capillary flow regime, both models produce the same results. With an increase in the external pressure applied to the liquid at the entrance to the junction, the results obtained by these models start diverging from each other. This demonstrates that, at large external pressures, the flows in the junction depart from the creeping flow approximation that was expected. We also showed that the meniscus arrest time in asymmetric Y-shape junction [18] decreases with an increase in the external pressure; when this pressure becomes large enough, the meniscus arrest disappears. We demonstrated that even in the capillary regime, the meniscus arrest may not appear if the entrance cross-section area of capillary 3 in Fig 1 is too small with respect to the entrance area in capillary 2.

In this article, we also investigated the applicability of the fully developed flow approximation in modeling capillary flows.

**Acknowledgements**

The author would like to express his sincere gratitude to his colleagues Dan Barnett, Paul Hoisington, and David Pekker for their kind support and helpful discussions. Special thanks to Chris Menzel who initiated this research.

**Appendix A**

In this Appendix, we derive Eq. (3.7). Taking into account the symmetry of flow, we integrate Eqs. (3.1) over $\int_0^Y \int_0^X dxdy$, Fig.4. Taking into account Eq. (3.4), the integration of the LHS of Eq. (3.1) yields:

$$\text{LHS: } \int_0^Y \int_0^X \frac{\partial v_z}{\partial t} dxdy + \int_0^Y \int_0^X \left( v_x \frac{\partial v_z}{\partial x} + v_y \frac{\partial v_z}{\partial y} + v_z \frac{\partial v_z}{\partial z} \right) dxdy =$$

$$= \frac{\partial}{\partial t} \left( \int_0^Y \int_0^X v_z dxdy \right) + \int_0^Y \int_0^X \left( \frac{\partial (v_z v_x)}{\partial x} - v_z \frac{\partial v_x}{\partial x} + \frac{\partial (v_z v_y)}{\partial y} - v_z \frac{\partial v_y}{\partial y} + v_z \frac{\partial v_z}{\partial z} \right) dxdy =$$

$$= \frac{\partial}{\partial t} \left( \int_0^Y \int_0^X v_z dxdy \right) + \int_0^Y \int_0^X \left( \frac{\partial (v_z v_x)}{\partial x} + \frac{\partial (v_z v_y)}{\partial y} + 2v_z \frac{\partial v_z}{\partial z} \right) dxdy =$$

$$= \frac{\partial}{\partial t} \left( \int_0^Y \int_0^X v_z dxdy \right) + \int_0^Y \int_0^X \left( \frac{\partial (v_z v_x)}{\partial x} + \frac{\partial (v_z v_y)}{\partial y} + \frac{\partial v_z^2}{\partial z} \right) dxdy =$$

$$= \frac{\partial}{\partial t}\left(\int_0^Y \int_0^X v_z dx dy\right) + \int_0^Y (v_z v_x)_0^X dy + \int_0^X (v_z v_y)_0^Y dx +$$

$$+ \frac{\partial}{\partial z}\left(\int_0^Y \int_0^X v_z^2 dx dy\right) - \frac{dX}{dz}\int_0^Y (v_z^2)_X dy - \frac{dY}{dz}\int_0^X (v_z^2)_Y dx =$$

$$= \frac{\partial}{\partial t}\left(\int_0^Y \int_0^X v_z dx dy\right) + \frac{\partial}{\partial z}\left(\int_0^Y \int_0^X v_z^2 dx dy\right) \qquad (A.1)$$

The integration the RHS of Eq. (3.1) yields:

$$\text{RHS: } \int_0^Y \int_0^X \left(-\frac{1}{\rho}\cdot\frac{\partial P}{\partial z} + \frac{\mu}{\rho}\left(\frac{\partial^2 v_z}{\partial x^2} + \frac{\partial^2 v_z}{\partial y^2} + \frac{\partial^2 v_z}{\partial z^2}\right)\right) dx dy =$$

$$= -\frac{XY}{\rho}\frac{\partial P}{\partial z} + \frac{\mu}{\rho}\left(\begin{array}{c}\int_0^Y \left(\frac{\partial v_z}{\partial x}\right)_X dy + \int_0^X \left(\frac{\partial v_z}{\partial y}\right)_Y dx + \frac{\partial}{\partial z}\left(\int_0^Y \int_0^X \frac{\partial v_z}{\partial z} dx dy\right) \\ -\frac{dX}{dz}\int_0^Y \left(\frac{\partial v_z}{\partial z}\right)_X dy - \frac{dY}{dz}\int_0^X \left(\frac{\partial v_z}{\partial z}\right)_Y dx\end{array}\right) =$$

$$= -\frac{XY}{\rho}\frac{\partial P}{\partial z} + \frac{\mu}{\rho}\left(\begin{array}{c}\int_0^Y \left(\frac{\partial v_z}{\partial x}\right)_X dy + \int_0^X \left(\frac{\partial v_z}{\partial y}\right)_Y dx - \frac{\partial}{\partial z}\left(\int_0^Y \int_0^X \left(\frac{\partial v_x}{\partial x} + \frac{\partial v_y}{\partial y}\right) dx dy\right) \\ -\frac{dX}{dz}\int_0^Y \left(\frac{\partial v_z}{\partial z}\right)_X dy - \frac{dY}{dz}\int_0^X \left(\frac{\partial v_z}{\partial z}\right)_Y dx\end{array}\right) =$$

$$= -\frac{XY}{\rho}\frac{\partial P}{\partial z} + \frac{\mu}{\rho}\left(\begin{array}{c}\int_0^Y \left(\frac{\partial v_z}{\partial x}\right)_X dy + \int_0^X \left(\frac{\partial v_z}{\partial y}\right)_Y dx - \frac{\partial}{\partial z}\left(\int_0^Y (v_x)_X dy\right) - \frac{\partial}{\partial z}\left(\int_0^X (v_y)_Y dx\right) \\ -\frac{dX}{dz}\int_0^Y \left(\frac{\partial v_z}{\partial z}\right)_X dy - \frac{dY}{dz}\int_0^X \left(\frac{\partial v_z}{\partial z}\right)_Y dx\end{array}\right) =$$

$$= -\frac{XY}{\rho}\frac{\partial P}{\partial z} + \frac{\mu}{\rho}\left(\int_0^Y \left(\frac{\partial v_z}{\partial x}\right)_X dy + \int_0^X \left(\frac{\partial v_z}{\partial y}\right)_Y dx - \frac{dX}{dz}\int_0^Y \left(\frac{\partial v_z}{\partial z}\right)_X dy - \frac{dY}{dz}\int_0^X \left(\frac{\partial v_z}{\partial z}\right)_Y dx\right) \quad (A.2)$$

Equating Eqs. (A.1) and Eq. (A.2), we obtain an equation for the z-momentum of the jet integrated over the cross-section of the channel:

$$\frac{\partial}{\partial t}\left(\int_0^Y \int_0^X v_z dx dy\right) + \frac{\partial}{\partial z}\left(\int_0^Y \int_0^X v_z^2 dx dy\right) =$$

$$= -\frac{XY}{\rho}\frac{\partial P}{\partial z} + \frac{\mu}{\rho}\left(\int_0^Y \left(\frac{\partial v_z}{\partial x}\right)_X dy + \int_0^X \left(\frac{\partial v_z}{\partial y}\right)_Y dx - \frac{dX}{dz}\int_0^Y \left(\frac{\partial v_z}{\partial z}\right)_X dy - \frac{dY}{dz}\int_0^X \left(\frac{\partial v_z}{\partial z}\right)_Y dx\right) \rightarrow$$

$$\rightarrow \frac{\rho}{XY}\frac{\partial}{\partial t}\left(\int_0^Y \int_0^X v_z dx dy\right) + \frac{1}{XY}\frac{\partial}{\partial z}\left(\rho \int_0^Y \int_0^X v_z^2 dx dy\right) =$$

$$= -\frac{\partial P}{\partial z} + \frac{\mu}{XY}\left(\int_0^Y \left(\frac{\partial v_z}{\partial x}\right)_X dy + \int_0^X \left(\frac{\partial v_z}{\partial y}\right)_Y dx - \frac{dX}{dz}\int_0^Y \left(\frac{\partial v_z}{\partial z}\right)_X dy - \frac{dY}{dz}\int_0^X \left(\frac{\partial v_z}{\partial z}\right)_Y dx\right) \rightarrow$$

$$\rightarrow \frac{\rho}{XY}\frac{\partial}{\partial t}\left(\int_0^Y \int_0^X v_z dx dy\right) + \frac{\partial}{\partial z}\left(\frac{\rho}{XY}\int_0^Y \int_0^X v_z^2 dx dy\right) - \frac{\rho}{X^2 Y}\frac{dX}{dz}\int_0^Y \int_0^X v_z^2 dx dy -$$

$$- \frac{\rho}{Y^2 X}\frac{dY}{dz}\int_0^Y \int_0^X v_z^2 dx dy = -\frac{\partial P}{\partial z} +$$

$$+\frac{\mu}{XY}\left(\int_0^Y \left(\frac{\partial v_z}{\partial x}\right)_X dy + \int_0^X \left(\frac{\partial v_z}{\partial y}\right)_Y dx - \frac{dX}{dz}\int_0^Y \left(\frac{\partial v_z}{\partial z}\right)_X dy - \frac{dY}{dz}\int_0^X \left(\frac{\partial v_z}{\partial z}\right)_Y dx\right) \quad (A.3)$$

Taking into that the flow, $\Phi = \int_0^Y \int_0^X v_z dx dy$, is independent of z, the integration of Eq. (A.3) over z from $z = 0$ to $z = Z$, Fig. 4, yields the following equation for the total z-momentum of the jet:

$$\left(\rho \frac{\partial}{\partial t}\left(\int_0^Y \int_0^X v_z dx dy\right)\right)\int_0^Z \frac{dz}{XY} - \int_0^Z \frac{\rho}{X^2 Y}\frac{dX}{dz}\left(\int_0^Y \int_0^X v_z^2 dx dy\right) dz - \int_0^Z \frac{\rho}{Y^2 X}\frac{dY}{dz}\left(\int_0^Y \int_0^X v_z^2 dx dy\right) dz =$$

$$= -\left(P + \frac{\rho}{XY}\int_0^Y \int_0^X v_z^2 dx dy\right)_0^Z +$$

$$+ \int_0^Z \frac{\mu}{XY}\left(\int_0^Y \left(\frac{\partial v_z}{\partial x}\right)_X dy + \int_0^X \left(\frac{\partial v_z}{\partial y}\right)_Y dx - \frac{dX}{dz}\int_0^Y \left(\frac{\partial v_z}{\partial z}\right)_X dy - \frac{dY}{dz}\int_0^X \left(\frac{\partial v_z}{\partial z}\right)_Y dx\right) dz \rightarrow$$

$$\rightarrow \left(\rho \frac{\partial}{\partial t}\left(\int_0^Y \int_0^X v_z dx dy\right)\right)\int_0^Z \frac{dz}{XY} - \int_0^Z \left(\frac{\rho}{Y^2 X}\frac{dY}{dz} + \frac{\rho}{X^2 Y}\frac{dX}{dz}\right)\left(\int_0^Y \int_0^X v_z^2 dx dy\right) dz =$$

$$= -\left(P + \frac{\rho}{XY}\int_0^Y \int_0^X v_z^2 dx dy\right)_0^Z +$$

$$+ \int_0^Z \frac{\mu}{XY}\left(\int_0^Y \left(\frac{\partial v_z}{\partial x}\right)_X dy + \int_0^X \left(\frac{\partial v_z}{\partial y}\right)_Y dx - \frac{dX}{dz}\int_0^Y \left(\frac{\partial v_z}{\partial z}\right)_X dy - \frac{dY}{dz}\int_0^X \left(\frac{\partial v_z}{\partial z}\right)_Y dx\right) dz \rightarrow$$

$$\rightarrow \left(\rho \frac{\partial}{\partial t}\left(\int_0^Y \int_0^X v_z dx dy\right)\right)\int_0^Z \frac{dz}{XY} + \int_0^Z \left(\frac{\rho}{Y^3 X^2}\frac{dY}{dz} + \frac{\rho}{X^3 Y^2}\frac{dX}{dz}\right)\left(XY \int_0^Y \int_0^X v_z^2 dx dy\right) dz =$$

$$= -\left(P + \frac{\rho}{XY}\int_0^Y \int_0^X v_z^2 dx dy\right)_0^Z +$$

$$+ \int_0^Z \frac{\mu}{XY}\left(\int_0^Y \left(\frac{\partial v_z}{\partial x}\right)_X dy + \int_0^X \left(\frac{\partial v_z}{\partial y}\right)_Y dx - \frac{dX}{dz}\int_0^Y \left(\frac{\partial v_z}{\partial z}\right)_X dy - \frac{dY}{dz}\int_0^X \left(\frac{\partial v_z}{\partial z}\right)_Y dx\right) dz \rightarrow$$

$$\rightarrow \left(\rho \frac{\partial}{\partial t}\left(\int_0^Y \int_0^X v_z dx dy\right)\right)\int_0^Z \frac{dz}{XY} - \int_0^Z \left(\frac{\rho}{2}\frac{d}{dz}\left(\frac{1}{X^2 Y^2}\right)\right)\left(XY \int_0^Y \int_0^X v_z^2 dx dy\right) dz =$$

$$= -\left(P + \frac{\rho}{XY}\int_0^Y \int_0^X v_z^2 dx dy\right)_0^Z +$$

$$+ \int_0^L \frac{\mu}{XY}\left(\int_0^Y \left(\frac{\partial v_z}{\partial x}\right)_X dy + \int_0^X \left(\frac{\partial v_z}{\partial y}\right)_Y dx - \frac{dX}{dz}\int_0^Y \left(\frac{\partial v_z}{\partial z}\right)_X dy - \frac{dY}{dz}\int_0^X \left(\frac{\partial v_z}{\partial z}\right)_Y dx\right) dz \rightarrow$$

$$\rightarrow \left(\rho \frac{\partial}{\partial t}\left(\int_0^Y \int_0^X v_z dx dy\right)\right)\int_0^L \frac{dz}{XY} -$$

$$- \int_0^Z \frac{\mu}{XY}\left(\int_0^Y \left(\frac{\partial v_z}{\partial x}\right)_X dy + \int_0^X \left(\frac{\partial v_z}{\partial y}\right)_Y dx - \frac{dX}{dz}\int_0^Y \left(\frac{\partial v_z}{\partial z}\right)_X dy - \frac{dY}{dz}\int_0^X \left(\frac{\partial v_z}{\partial z}\right)_Y dx\right) dz =$$

$$= P_0(t) - ST + \left(\frac{\rho}{2XY}\int_0^Y \int_0^X v_z^2 dx dy\right)_{z=0} - \left(\frac{\rho}{2XY}\int_0^Y \int_0^X v_z^2 dx dy\right)_{z=Z} -$$

$$-\frac{\rho}{2}\int_0^Z \left(\frac{1}{X^2Y^2}\frac{\partial}{\partial z}\left(XY\int_0^Y\int_0^X v_Z^2 dx dy\right)\right)dz \tag{A.4}$$

In Eq. (A.4), we have substitute $(P)_0$ as $P_0(t)$, $(P)^Z$ as $ST$.

## Appendix B

In this Appendix, we derive Eq. (3.12). We begin by presenting $v_z$ from Eq. (3.10) in the following form:

$$v_z(x,y) = \frac{\pi\Phi}{8XY}\frac{\sum_{i=0}^{i=\infty}\frac{(-1)^i}{(2i+1)^3}f_i(x,y)}{\sum_{i=0}^{i=\infty}\frac{1}{(2i+1)^4}\left(1-\frac{2X}{(2i+1)\pi Y}tgh\left(\frac{(2i+1)\pi Y}{2X}\right)\right)} \tag{B.1}$$

where

$$f_i(x,y) = \cos\left(\frac{(2i+1)\pi}{2}\frac{x}{X}\right)\cos\left(1-\frac{\cosh\left(\frac{(2i+1)\pi}{2}\frac{y}{X}\right)}{\cosh\left(\frac{(2i+1)\pi Y}{2X}\right)}\right) \tag{B.2}$$

As one can see, the base functions $f_i(x,y)$ are orthogonal to each other,

$$\int_0^X\int_0^Y f_i(x,y)f_j(x,y)dxdy = 0 \quad i \neq j \tag{B.3}$$

Using Eqs. (B.1), (B.2) and $\xi$ from Eq. (3.17), we derive a set of expressions presented in Eq. (3.12):

$$\int_0^Y\int_0^X v_Z^2 dxdy =$$

$$= \left(\frac{\pi\Phi}{8XY}\right)^2 \frac{\left(\sum_{i=0}^{i=\infty}\left(\frac{(-1)^i}{(2i+1)^3}\right)^2\left(\int_0^X\left(\cos\left(\frac{(2i+1)\pi}{2}\frac{x}{X}\right)\right)^2 dx\right)\left(\int_0^Y\left(1-\frac{\cosh\left(\frac{(2i+1)\pi}{2}\frac{y}{X}\right)}{\cosh\left(\frac{(2i+1)\pi Y}{2X}\right)}\right)^2 dy\right)\right)}{\left(\sum_{i=0}^{i=\infty}\frac{1}{(2i+1)^4}\left(1-\frac{2X}{(2i+1)\pi Y}tgh\left(\frac{(2i+1)\pi Y}{2X}\right)\right)\right)^2} =$$

$$= \left(\frac{\pi\Phi}{8XY}\right)^2 \frac{\left(\sum_{i=0}^{i=\infty}\frac{1}{(2i+1)^6}\left(\frac{1}{2}X\right)\left(\int_0^Y\left(1-\frac{\cosh\left(\frac{(2i+1)\pi\xi}{2}\frac{y}{Y}\right)}{\cosh\left(\frac{(2i+1)\pi\xi}{2}\right)}\right)^2 dy\right)\right)}{\left(\sum_{i=0}^{i=\infty}\frac{1}{(2i+1)^4}\left(1-\frac{2}{(2i+1)\pi\xi}tgh\left(\frac{(2i+1)\pi Y}{2X}\right)\right)\right)^2} =$$

$$= \frac{\pi^2}{128}\frac{\Phi^2}{XY}\frac{\left(\sum_{i=0}^{i=\infty}\frac{1}{(2i+1)^6}\left(\int_0^1\left(1-\frac{\cosh\left(\frac{(2i+1)\pi\xi}{2}\hat{y}\right)}{\cosh\left(\frac{(2i+1)\pi Y}{2X}\right)}\right)^2 d\hat{y}\right)\right)}{\left(\sum_{i=0}^{i=\infty}\frac{1}{(2i+1)^4}\left(1-\frac{2}{(2i+1)\pi\xi}tgh\left(\frac{(2i+1)\pi\xi}{2}\right)\right)\right)^2} \rightarrow$$

$$\rightarrow \int_0^Y\int_0^X v_Z^2 dxdy = \frac{\pi^2}{128}\frac{\Phi^2}{XY}U(\xi) \tag{B.4}$$

$$\left(\frac{\partial v_z}{\partial x}\right)_{x=X} = \frac{\pi\Phi}{8XY} \frac{\sum_{i=0}^{i=\infty}\frac{(-1)^i}{(2i+1)^3}\left(\frac{\partial}{\partial x}\left(cos\left(\frac{(2i+1)\pi}{2}\frac{x}{X}\right)\right)\right)_{x=X}\left(1-\frac{cosh\left(\frac{(2i+1)\pi\,y}{2\,X}\right)}{cosh\left(\frac{(2i+1)\pi\xi}{2}\right)}\right)^2}{\sum_{i=0}^{i=\infty}\frac{1}{(2i+1)^4}\left(1-\frac{2X}{(2i+1)\pi Y}tgh\left(\frac{(2i+1)\pi Y}{2X}\right)\right)} =$$

$$= \frac{\pi\Phi}{8XY}\ \frac{\sum_{i=0}^{i=\infty}\frac{(-1)^i}{(2i+1)^3}\frac{(2i+1)\pi}{2X}\frac{dX}{dz}\left(sin\left(i\pi+\frac{\pi}{2}\right)\right)\left(1-\frac{cosh\left(\frac{(2i+1)\pi y}{2X}\right)}{cosh\left(\frac{(2i+1)\pi\xi}{2}\right)}\right)}{\sum_{i=0}^{i=\infty}\frac{1}{(2i+1)^4}\left(1-\frac{2}{(2i+1)\pi\xi}tgh\left(\frac{(2i+1)\pi\xi}{2}\right)\right)} =$$

$$= -\frac{\pi^2\Phi}{16X^2Y}\frac{\sum_{i=0}^{i=\infty}\frac{1}{(2i+1)^2}\left(1-\frac{cosh\left(\frac{(2i+1)\pi y}{2X}\right)}{cosh\left(\frac{(2i+1)\pi\xi}{2}\right)}\right)}{\sum_{i=0}^{i=\infty}\frac{1}{(2i+1)^4}\left(1-\frac{2}{(2i+1)\pi\xi}tgh\left(\frac{(2i+1)\pi\xi}{2}\right)\right)} \rightarrow$$

$$\rightarrow \left(\frac{\partial v_z}{\partial x}\right)_{x=X} = -\frac{\pi^2\Phi}{16X^2Y}\frac{\sum_{i=0}^{i=\infty}\frac{1}{(2i+1)^2}\left(1-\frac{cosh\left(\frac{(2i+1)\pi y}{2X}\right)}{cosh\left(\frac{(2i+1)\pi\xi}{2}\right)}\right)}{\sum_{i=0}^{i=\infty}\frac{1}{(2i+1)^4}\left(1-\frac{2}{(2i+1)\pi\xi}tgh\left(\frac{(2i+1)\pi\xi}{2}\right)\right)} \qquad (B.5)$$

$$\left(\frac{\partial v_z}{\partial z}\right)_{x=X} = \frac{\pi\Phi}{8XY}\frac{\sum_{i=0}^{i=\infty}\frac{(-1)^i}{(2i+1)^3}\left(\frac{\partial}{\partial z}\left(cos\left(\frac{(2i+1)\pi x}{2X}\right)\right)\right)_{x=X}\left(1-\frac{cosh\left(\frac{(2i+1)\pi y}{2X}\right)}{cosh\left(\frac{(2i+1)\pi\xi}{2}\right)}\right)}{\sum_{i=0}^{i=\infty}\frac{1}{(2i+1)^4}\left(1-\frac{2X}{(2i+1)\pi Y}tgh\left(\frac{(2i+1)\pi Y}{2X}\right)\right)} =$$

$$= \frac{\pi\Phi}{8XY}\ \frac{\sum_{i=0}^{i=\infty}\frac{(-1)^i}{(2i+1)^3}\frac{(2i+1)\pi}{2X}\frac{dX}{dz}\left(sin\left(i\pi+\frac{\pi}{2}\right)\right)\left(1-\frac{cosh\left(\frac{(2i+1)\pi y}{2X}\right)}{cosh\left(\frac{(2i+1)\pi\xi}{2}\right)}\right)}{\sum_{i=0}^{i=\infty}\frac{1}{(2i+1)^4}\left(1-\frac{2}{(2i+1)\pi\xi}tgh\left(\frac{(2i+1)\pi\xi}{2}\right)\right)} =$$

$$= \frac{\pi^2\Phi}{16X^2Y}\frac{dX}{dz}\frac{\sum_{i=0}^{i=\infty}\frac{1}{(2i+1)^2}\left(1-\frac{cosh\left(\frac{(2i+1)\pi y}{2X}\right)}{cosh\left(\frac{(2i+1)\pi\xi}{2}\right)}\right)}{\sum_{i=0}^{i=\infty}\frac{1}{(2i+1)^4}\left(1-\frac{2}{(2i+1)\pi\xi}tgh\left(\frac{(2i+1)\pi\xi}{2}\right)\right)}$$

$$\rightarrow \left(\frac{\partial v_z}{\partial z}\right)_{x=X} = \frac{\pi^2\Phi}{16X^2Y}\frac{dX}{dz}\frac{\sum_{i=0}^{i=\infty}\frac{1}{(2i+1)^2}\left(1-\frac{cosh\left(\frac{(2i+1)\pi y}{2X}\right)}{cosh\left(\frac{(2i+1)\pi\xi}{2}\right)}\right)}{\sum_{i=0}^{i=\infty}\frac{1}{(2i+1)^4}\left(1-\frac{2}{(2i+1)\pi\xi}tgh\left(\frac{(2i+1)\pi\xi}{2}\right)\right)} \qquad (B.6)$$

$$\left(\frac{\partial v_z}{\partial y}\right)_{y=Y} = \frac{\pi\Phi}{8XY}\frac{\sum_{i=0}^{i=\infty}\frac{(-1)^i}{(2i+1)^3}cos\left(\frac{(2i+1)\pi x}{2X}\right)\left(\frac{\partial}{\partial y}\left(1-\frac{cosh\left(\frac{(2i+1)\pi y}{2X}\right)}{cosh\left(\frac{(2i+1)\pi\xi}{2}\right)}\right)\right)_{y=Y}}{\sum_{i=0}^{i=\infty}\frac{1}{(2i+1)^4}\left(1-\frac{2X}{(2i+1)\pi Y}tgh\left(\frac{(2i+1)\pi Y}{2X}\right)\right)} =$$

$$= -\frac{\pi\Phi}{8XY} \frac{\sum_{i=0}^{i=\infty} \frac{(-1)^i}{(2i+1)^3} \cos\left(\frac{(2i+1)\pi x}{2X}\right) \frac{(2i+1)\pi}{2X} \frac{\sinh\left(\frac{(2i+1)\pi Y}{2X}\right)}{\cosh\left(\frac{(2i+1)\pi \xi}{2}\right)}}{\sum_{i=0}^{i=\infty} \frac{1}{(2i+1)^4}\left(1 - \frac{2}{(2i+1)\pi\xi}tgh\left(\frac{(2i+1)\pi\xi}{2}\right)\right)} =$$

$$= -\frac{\pi^2\Phi}{16X^2Y} \frac{\sum_{i=0}^{i=\infty} \frac{(-1)^i}{(2i+1)^2} \cos\left(\frac{(2i+1)\pi x}{2X}\right) tgh\left(\frac{(2i+1)\pi\xi}{2}\right)}{\sum_{i=0}^{i=\infty} \frac{1}{(2i+1)^4}\left(1 - \frac{2}{(2i+1)\pi\xi}tgh\left(\frac{(2i+1)\pi\xi}{2}\right)\right)} \quad \rightarrow$$

$$\rightarrow \left(\frac{\partial v_z}{\partial y}\right)_{y=Y} = -\frac{\pi^2\Phi}{16X^2Y} \frac{\sum_{i=0}^{i=\infty} \frac{(-1)^i}{(2i+1)^2} \cos\left(\frac{(2i+1)\pi x}{2X}\right) tgh\left(\frac{(2i+1)\pi\xi}{2}\right)}{\sum_{i=0}^{i=\infty} \frac{1}{(2i+1)^4}\left(1 - \frac{2}{(2i+1)\pi\xi}tgh\left(\frac{(2i+1)\pi\xi}{2}\right)\right)} \quad (B.7)$$

$$\left(\frac{\partial v_z}{\partial z}\right)_{y=Y} = \frac{\pi\Phi}{8XY} \frac{\sum_{i=0}^{i=\infty} \frac{(-1)^i}{(2i+1)^3} \cos\left(\frac{(2i+1)\pi x}{2X}\right) \left(\frac{\partial}{\partial Z}\left(1 - \frac{\cosh\left(\frac{(2i+1)\pi y}{2X}\right)}{\cosh\left(\frac{(2i+1)\pi Y}{2X}\right)}\right)\right)_{y=Y}}{\sum_{i=0}^{i=\infty} \frac{1}{(2i+1)^4}\left(1 - \frac{2X}{(2i+1)\pi Y}tgh\left(\frac{(2i+1)\pi Y}{2X}\right)\right)} \quad (B.8)$$

Let us calculate the $\frac{\partial}{\partial Z}$ term in Eq. (B.8)

$$\left(\frac{\partial}{\partial Z}\left(1 - \frac{\cosh\left(\frac{(2i+1)\pi y}{2X}\right)}{\cosh\left(\frac{(2i+1)\pi Y}{2X}\right)}\right)\right)_{y=Y} =$$

$$= \frac{\sinh\left(\frac{(2i+1)\pi Y}{2X}\right)}{\cosh\left(\frac{(2i+1)\pi Y}{2X}\right)} \frac{(2i+1)\pi Y}{2X^2}\frac{dX}{dZ} + \frac{\cosh\left(\frac{(2i+1)\pi Y}{2X}\right)}{\left(\cosh\left(\frac{(2i+1)\pi Y}{2X}\right)\right)^2} \sinh\left(\frac{(2i+1)\pi Y}{2X}\right) \left(\frac{(2i+1)\pi}{2}\right)\left(-\frac{Y}{X^2}\frac{dX}{dZ} + \frac{1}{X}\frac{dY}{dZ}\right) =$$

$$= \frac{(2i+1)\pi}{2}\left(\frac{Y}{X^2}\frac{dX}{dZ}\frac{\sinh\left(\frac{(2i+1)\pi Y}{2X}\right)}{\cosh\left(\frac{(2i+1)\pi Y}{2X}\right)} + \frac{\sinh\left(\frac{(2i+1)\pi Y}{2X}\right)}{\cosh\left(\frac{(2i+1)\pi Y}{2X}\right)}\left(-\frac{Y}{X^2}\frac{dX}{dZ} + \frac{1}{X}\frac{dY}{dZ}\right)\right) =$$

$$= \frac{(2i+1)\pi}{2X}\frac{dY}{dZ}\tanh\left(\frac{(2i+1)\pi Y}{2X}\right) = \frac{(2i+1)\pi}{2X}\frac{dY}{dZ}tgh\left(\frac{(2i+1)\pi\xi}{2}\right) \quad (B.9)$$

Substituting Eq. (B.9) into Eq. (B.8) yields

$$\left(\frac{\partial v_z}{\partial z}\right)_{y=Y} = \frac{\pi^2\Phi}{16X^2Y}\frac{dY}{dZ} \frac{\sum_{i=0}^{i=\infty} \frac{(-1)^i}{(2i+1)^2} \cos\left(\frac{(2i+1)\pi x}{2X}\right) tgh\left(\frac{(2i+1)\pi\xi}{2}\right)}{\sum_{i=0}^{i=\infty} \frac{1}{(2i+1)^4}\left(1 - \frac{2}{(2i+1)\pi\xi}tgh\left(\frac{(2i+1)\pi\xi}{2}\right)\right)} \quad (B.10)$$

Substituting Eq. (B5) and (B.6) into $\int_0^Y \left(\frac{\partial v_z}{\partial x}\right)_{x=X} dy - \frac{dX}{dz}\int_0^Y \left(\frac{\partial v_z}{\partial z}\right)_{x=X} dy$ and Eqs. (B.7) and (B.10) into

$\int_0^X \left(\frac{\partial v_z}{\partial y}\right)_{y=Y} dx - \frac{dY}{dz}\int_0^X \left(\frac{\partial v_z}{\partial z}\right)_{y=Y} dx$ yields:

$$\int_0^Y \left(\frac{\partial v_z}{\partial x}\right)_{x=X} dy - \frac{dX}{dz}\int_0^Y \left(\frac{\partial v_z}{\partial z}\right)_{x=X} dy =$$

$$= \int_0^Y \left( \begin{array}{c} -\frac{\pi^2 \Phi}{16X^2 Y} \frac{\sum_{i=0}^{i=\infty}\frac{1}{(2i+1)^2}\left(1-\frac{\cosh\left(\frac{(2i+1)\pi y}{2X}\right)}{\cosh\left(\frac{(2i+1)\pi \xi}{2}\right)}\right)}{\sum_{i=0}^{i=\infty}\frac{1}{(2i+1)^4}\left(1-\frac{2}{(2i+1)\pi\xi}tgh\left(\frac{(2i+1)\pi\xi}{2}\right)\right)} - \\ -\frac{\pi^2 \Phi}{16X^2 Y}\left(\frac{dX}{dz}\right)^2 \frac{\sum_{i=0}^{i=\infty}\frac{1}{(2i+1)^2}\left(1-\frac{\cosh\left(\frac{(2i+1)\pi y}{2X}\right)}{\cosh\left(\frac{(2i+1)\pi \xi}{2}\right)}\right)}{\sum_{i=0}^{i=\infty}\frac{1}{(2i+1)^4}\left(1-\frac{2}{(2i+1)\pi\xi}tgh\left(\frac{(2i+1)\pi\xi}{2}\right)\right)} \end{array} \right) dy =$$

$$= -\frac{\pi^2 \Phi}{16X^2 Y}\left(1+\left(\frac{dX}{dz}\right)^2\right) \frac{\int_0^Y \left(\sum_{i=0}^{i=\infty}\frac{1}{(2i+1)^2}\left(1-\frac{\cosh\left(\frac{(2i+1)\pi y}{2X}\right)}{\cosh\left(\frac{(2i+1)\pi}{2\xi}\right)}\right)\right) dy}{\sum_{i=0}^{i=\infty}\frac{1}{(2i+1)^4}\left(1-\frac{2}{(2i+1)\pi\xi}tgh\left(\frac{(2i+1)\pi\xi}{2}\right)\right)} =$$

$$= -\frac{\pi^2 \Phi}{16X^2 Y}\left(1+\left(\frac{dX}{dz}\right)^2\right) \frac{\sum_{i=0}^{i=\infty}\frac{1}{(2i+1)^2}\left(Y-\frac{2X}{(2i+1)\pi}tgh\left(\frac{(2i+1)\pi Y}{2X}\right)\right)}{\sum_{i=0}^{i=\infty}\frac{1}{(2i+1)^4}\left(1-\frac{2}{(2i+1)\pi\xi}tgh\left(\frac{(2i+1)\pi\xi}{2}\right)\right)} \rightarrow$$

$$\rightarrow \int_0^Y \left(\frac{\partial v_z}{\partial x}\right)_{x=X} dy - \frac{dX}{dz}\int_0^Y \left(\frac{\partial v_z}{\partial z}\right)_{x=X} dy =$$

$$= -\frac{\pi^2 \Phi}{16XY}\left(1+\left(\frac{dX}{dz}\right)^2\right) \frac{\sum_{i=0}^{i=\infty}\frac{1}{(2i+1)^2}\left(\xi-\frac{2}{(2i+1)\pi}tgh\left(\frac{(2i+1)\pi\xi}{2}\right)\right)}{\sum_{i=0}^{i=\infty}\frac{1}{(2i+1)^4}\left(1-\frac{2}{(2i+1)\pi\xi}tgh\left(\frac{(2i+1)\pi\xi}{2}\right)\right)} \quad (B.11)$$

$$\int_0^X \left(\frac{\partial v_z}{\partial y}\right)_{y=Y} dx - \frac{dY}{dz}\int_0^X \left(\frac{\partial v_z}{\partial z}\right)_{y=Y} dx =$$

$$= \int_0^X \left( \begin{array}{c} -\frac{\pi^2 \Phi}{16X^2 Y} \frac{\sum_{i=0}^{i=\infty}\frac{(-1)^i}{(2i+1)^2}\cos\left(\frac{(2i+1)\pi x}{2X}\right) tgh\left(\frac{(2i+1)\pi\xi}{2}\right)}{\sum_{i=0}^{i=\infty}\frac{1}{(2i+1)^4}\left(1-\frac{2}{(2i+1)\pi\xi}tgh\left(\frac{(2i+1)\pi\xi}{2}\right)\right)} - \\ -\frac{\pi^2 \Phi}{16X^2 Y}\left(\frac{dY}{dZ}\right)^2 \frac{\sum_{i=0}^{i=\infty}\frac{(-1)^i}{(2i+1)^2}\cos\left(\frac{(2i+1)\pi x}{2X}\right) tgh\left(\frac{(2i+1)\pi\xi}{2}\right)}{\sum_{i=0}^{i=\infty}\frac{1}{(2i+1)^4}\left(1-\frac{2}{(2i+1)\pi\xi}tgh\left(\frac{(2i+1)\pi\xi}{2}\right)\right)} \end{array} \right) dx =$$

$$= -\frac{\pi^2 \Phi}{16X^2 Y}\left(1+\left(\frac{dY}{dZ}\right)^2\right) \frac{\sum_{i=0}^{i=\infty}\frac{(-1)^i}{(2i+1)^2}\left(\int_0^X \cos\left(\frac{(2i+1)\pi x}{2X}\right) dx\right) tgh\left(\frac{(2i+1)\pi\xi}{2}\right)}{\sum_{i=0}^{i=\infty}\frac{1}{(2i+1)^4}\left(1-\frac{2}{(2i+1)\pi\xi}tgh\left(\frac{(2i+1)\pi\xi}{2}\right)\right)} =$$

$$= -\frac{\pi^2 \Phi}{16X^2 Y}\left(1+\left(\frac{dY}{dZ}\right)^2\right) \frac{\sum_{i=0}^{i=\infty}\frac{(-1)^i}{(2i+1)^2}\frac{2X}{(2i+1)\pi}\sin\left(\frac{(2i+1)\pi}{2}\right) tgh\left(\frac{(2i+1)\pi\xi}{2}\right)}{\sum_{i=0}^{i=\infty}\frac{1}{(2i+1)^4}\left(1-\frac{2}{(2i+1)\pi\xi}tgh\left(\frac{(2i+1)\pi\xi}{2}\right)\right)} =$$

$$= -\frac{\pi\Phi}{8XY}\left(1+\left(\frac{dY}{dZ}\right)^2\right)\frac{\sum_{i=0}^{i=\infty}\frac{(-1)^i}{(2i+1)^3}(-1)^i tgh\left(\frac{(2i+1)\pi\xi}{2}\right)}{\sum_{i=0}^{i=\infty}\frac{1}{(2i+1)^4}\left(1-\frac{2}{(2i+1)\pi\xi}tgh\left(\frac{(2i+1)\pi\xi}{2}\right)\right)}=$$

$$= -\frac{\pi\Phi}{8XY}\left(1+\left(\frac{dY}{dZ}\right)^2\right)\frac{\sum_{i=0}^{i=\infty}\frac{1}{(2i+1)^3}tgh\left(\frac{(2i+1)\pi\xi}{2}\right)}{\sum_{i=0}^{i=\infty}\frac{1}{(2i+1)^4}\left(1-\frac{2}{(2i+1)\pi\xi}tgh\left(\frac{(2i+1)\pi\xi}{2}\right)\right)} \rightarrow$$

$$\rightarrow \int_0^X \left(\frac{\partial v_z}{\partial y}\right)_{y=Y} dx - \frac{dY}{dz}\int_0^X \left(\frac{\partial v_z}{\partial z}\right)_{y=Y} dx =$$

$$= -\frac{\pi\Phi}{8XY}\left(1+\left(\frac{dY}{dZ}\right)^2\right)\frac{\sum_{i=0}^{i=\infty}\frac{1}{(2i+1)^3}tgh\left(\frac{(2i+1)\pi\xi}{2}\right)}{\sum_{i=0}^{i=\infty}\frac{1}{(2i+1)^4}\left(1-\frac{2}{(2i+1)\pi\xi}tgh\left(\frac{(2i+1)\pi\xi}{2}\right)\right)} \tag{B.12}$$

Combining Eq. (B.12) and (B.12), we obtain

$$\int_0^Y \left(\frac{\partial v_z}{\partial x}\right)_{x=X} dy - \frac{dX}{dz}\int_0^Y \left(\frac{\partial v_z}{\partial z}\right)_{x=X} dy + \int_0^X \left(\frac{\partial v_z}{\partial y}\right)_{y=Y} dx - \frac{dY}{dz}\int_0^X \left(\frac{\partial v_z}{\partial z}\right)_{y=Y} dx =$$

$$= -\frac{\pi^2\Phi}{16XY}\left(1+\left(\frac{dX}{dz}\right)^2\right)\frac{\sum_{i=0}^{i=\infty}\frac{1}{(2i+1)^2}\left(\xi-\frac{2}{(2i+1)\pi}tgh\left(\frac{(2i+1)\pi\xi}{2}\right)\right)}{\sum_{i=0}^{i=\infty}\frac{1}{(2i+1)^4}\left(1-\frac{2}{(2i+1)\pi\xi}tgh\left(\frac{(2i+1)\pi\xi}{2}\right)\right)}-$$

$$-\frac{\pi\Phi}{8XY}\left(1+\left(\frac{dY}{dZ}\right)^2\right)\frac{\sum_{i=0}^{i=\infty}\frac{1}{(2i+1)^3}tgh\left(\frac{(2i+1)\pi\xi}{2}\right)}{\sum_{i=0}^{i=\infty}\frac{1}{(2i+1)^4}\left(1-\frac{2}{(2i+1)\pi\xi}tgh\left(\frac{(2i+1)\pi\xi}{2}\right)\right)} =$$

$$= -\frac{\pi^2\Phi}{16XY}\frac{\sum_{i=0}^{i=\infty}\frac{1}{(2i+1)^2}\left(\xi-\frac{2}{(2i+1)\pi}tgh\left(\frac{(2i+1)\pi}{2\xi}\right)+\frac{2}{(2i+1)\pi}tgh\left(\frac{(2i+1)\pi}{2\xi}\right)\right)}{\sum_{i=0}^{i=\infty}\frac{1}{(2i+1)^4}\left(1-\frac{2}{(2i+1)\pi\xi}tgh\left(\frac{(2i+1)\pi\xi}{2}\right)\right)}-$$

$$-\frac{\pi^2\Phi}{16XY}\left(\frac{dX}{dz}\right)^2\frac{\sum_{i=0}^{i=\infty}\frac{1}{(2i+1)^2}\left(\xi-\frac{2}{(2i+1)\pi}tgh\left(\frac{(2i+1)\pi\xi}{2}\right)\right)}{\sum_{i=0}^{i=\infty}\frac{1}{(2i+1)^4}\left(1-\frac{2}{(2i+1)\pi\xi}tgh\left(\frac{(2i+1)\pi\xi}{2}\right)\right)}-$$

$$-\frac{\pi^2\Phi}{16XY}\left(\frac{dY}{dZ}\right)^2\frac{\sum_{i=0}^{i=\infty}\frac{2}{\pi(2i+1)^3}tgh\left(\frac{(2i+1)\pi}{2\xi}\right)}{\sum_{i=0}^{i=\infty}\frac{1}{(2i+1)^4}\left(1-\frac{2}{(2i+1)\pi\xi}tgh\left(\frac{(2i+1)\pi}{2\xi}\right)\right)} =$$

$$= -\frac{\pi^2\Phi}{16XY}\frac{\sum_{i=0}^{i=\infty}\frac{\xi}{(2i+1)^2}}{\sum_{i=0}^{i=\infty}\frac{1}{(2i+1)^4}\left(1-\frac{2}{(2i+1)\pi\xi}tgh\left(\frac{(2i+1)\pi}{2\xi}\right)\right)}-$$

$$-\frac{\pi^2\Phi}{16XY}\left(\frac{dX}{dz}\right)^2\frac{\left(\sum_{i=0}^{i=\infty}\frac{\xi}{(2i+1)^2}\right)-\left(\sum_{i=0}^{i=\infty}\frac{2\xi}{\pi(2i+1)^3}tgh\left(\frac{(2i+1)\pi}{2\xi}\right)\right)}{\sum_{i=0}^{i=\infty}\frac{1}{(2i+1)^4}\left(1-\frac{2}{(2i+1)\pi\xi}tgh\left(\frac{(2i+1)\pi}{2\xi}\right)\right)}-$$

$$-\frac{\pi^2 \Phi}{16XY}\left(\frac{dY}{dZ}\right)^2 \frac{\sum_{i=0}^{i=\infty}\frac{2}{\pi(2i+1)^3}tgh\left(\frac{(2i+1)\pi}{2\xi}\right)}{\sum_{i=0}^{i=\infty}\frac{1}{(2i+1)^4}\left(1-\frac{2}{(2i+1)\pi\xi}tgh\left(\frac{(2i+1)\pi}{2\xi}\right)\right)} \rightarrow$$

$$\rightarrow \int_0^Y \left(\frac{\partial v_z}{\partial x}\right)_{x=X} dy - \frac{dX}{dz}\int_0^Y \left(\frac{\partial v_z}{\partial z}\right)_{x=X} dy + \int_0^X \left(\frac{\partial v_z}{\partial y}\right)_{y=Y} dx - \frac{dY}{dz}\int_0^X \left(\frac{\partial v_z}{\partial z}\right)_{y=Y} dx =$$

$$= -\frac{\pi^2 \Phi}{16XY}\left(F_{unity}(\xi) + \left(\frac{dX}{dz}\right)^2 F_{xx}(\xi) + \left(\frac{dY}{dZ}\right)^2 F_{yy}(\xi)\right) \tag{B.13}$$

Substituting Eqs. (3.11), (B.4) and (B.13) into Eq. (3.7), we obtain Eq. (3.12)

$$\frac{\rho}{4}\frac{\partial \Phi}{\partial t}\int_0^Z \frac{dz}{XY} + \frac{\pi^2 \mu \Phi}{16}\int_0^Z F_{unity}(\xi) + \left(\frac{dX}{dz}\right)^2 F_{xx}(\xi) + \left(\frac{dY}{dZ}\right)^2 F_{yy}(\xi)\frac{dz}{X^2Y^2} =$$

$$= P_0(t) - ST + \frac{\pi^2 \Phi^2 \rho}{256}\left(\frac{1}{X^2Y^2}U(\xi)\right)_{z=0} - \frac{\pi^2 \Phi^2 \rho}{256}\left(\frac{1}{X^2Y^2}U(\xi)\right)_{z=Z} -$$

$$-\frac{\pi^2 \rho \Phi^2}{256}\int_0^Z \left(\frac{1}{X^2Y^2}\frac{\partial}{\partial z}(U(\xi))\right)dz \rightarrow$$

$$\rightarrow \frac{\rho}{4}\frac{\partial \Phi}{\partial t}\int_0^Z \frac{dz}{XY} + \frac{\pi^2 \mu \Phi}{16}\int_0^Z F_{unity}(\xi) + \left(\frac{dX}{dz}\right)^2 F_{xx}(\xi) + \left(\frac{dY}{dZ}\right)^2 F_{yy}(\xi)\frac{dz}{X^2Y^2} =$$

$$= P_0(t) - ST + \frac{\pi^2 \Phi^2 \rho}{256}\left(\frac{1}{X^2Y^2}U(\xi)\right)_{z=0} - \frac{\pi^2 \Phi^2 \rho}{256}\left(\frac{1}{X^2Y^2}U(\xi)\right)_{z=Z} -$$

$$-\frac{\pi^2 \rho \Phi^2}{256}\int_0^Z \left(\frac{1}{X^2Y^2}\frac{d\xi}{dz}\frac{dU(\xi)}{d\xi}\right)dz \tag{B.14}$$

**Appendix C**

In the code, in the range of $1 \le \xi \le 5$, we use the following polynomial approximation functions for $U, \frac{dU}{d\xi}$, and $F_{unity}$:

$$U(\xi) = U_0(\xi) = 0.000305\xi^5 - 0.005266\xi^4 + 0.035064\xi^3 - 0.108017\xi^2 +$$
$$+0.122664\xi + 1.073108 \tag{C.1}$$

$$\frac{dU(\xi)}{d\xi} = dUd\xi_0(\xi) = 0.00023961\xi^6 - 0.0049973\xi^5 + 0.04301008\xi^4$$
$$-0.19598432\xi^3 + 0.4986699\xi^2 - 0.661616\xi + 0.32014735 \tag{C.2}$$

$$F_{unity}(\xi) = F_{unity-0}(\xi) = -0.004805\xi^5 + 0.084263\xi^4 - 0.586338\xi^3 + 2.048328\xi^2 -$$
$$-2.486629\xi + 3.815105 \tag{C.3}$$

Taking into account the property of functions $U, F_{unity}$ presented in Eq. (3.18) and the fact that

$$\frac{dU\left(\frac{1}{\xi}\right)}{d\xi} = -\frac{dU\left(\frac{1}{\xi}\right)}{d\left(\frac{1}{\xi}\right)} \frac{1}{\xi^2} \tag{C.4}$$

we obtain that, in the range of $0.2 \leq \xi \leq 1$, the approximate functions for $U, \frac{dU}{d\xi}, F_{unity}$ satisfy:

$$U(\xi) = U_0(\xi^{-1}) \quad \frac{dU(\xi)}{d\xi} = -\frac{dUd\xi_0(\xi^{-1})}{\xi^2} \quad F_{unity}(\xi) = F_{unity-0}(\xi^{-1}) \tag{C.5}$$

For $F_{xx}$ and $F_{yy}$, in the range of $0.2 \leq \xi \leq 5$, we use the following polynomial approximation functions:

$$F_{xx}(\xi) = F_{xx-0}(\xi) = 0.000688\xi^6 - 0.012683\xi^5 + 0.09463\xi^4 - 0.368021\xi^3 +$$
$$+0.800014\xi^2 + 0.252873\xi + 0.670972 \tag{C.6}$$

$$F_{yy}(\xi) = F_{xx-0}(\xi^{-1}) = 0.000688\xi^{-6} - 0.012683\xi^{-5} + 0.09463\xi^{-4} - 0.368021\xi^{-3} +$$
$$+0.800014\xi^{-2} + 0.252873\xi^{-1} + 0.670972 \tag{C.7}$$

In Eq. (C.7), we have used the relationship between $F_{xx}$ and $F_{yy}$ presented in Eq. (3.18).

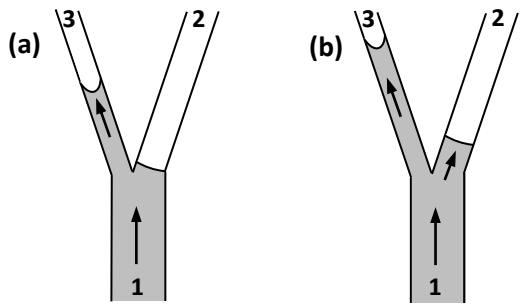

Fig. 1. Illustration of the meniscus arrest during capillary rise in an asymmetric Y-shape junction [18]; the radius of channel 3 is smaller than the radius of channel 2. (a) – Liquid is filling channel 3 while the meniscus (flow) in channel 2 is arrested; (b) – When the pressure in the junction becomes small enough the liquid starts filling channel 2 as well.

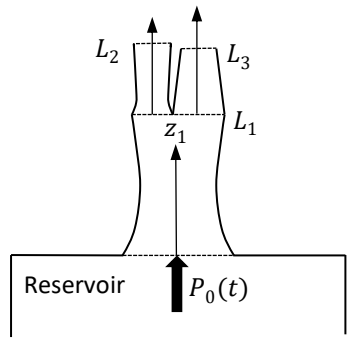

Fig. 2. A principal schema of our three-channel $Y$-shape capillary network model. The network consists of three tapered circular or rectangular channels. $L_1$, $L_2$, $L_3$ are the lengths of cannels 1, 2, and 3 respectively; $P_0$ is the external pressure applied to the liquid at the entrance to the network.

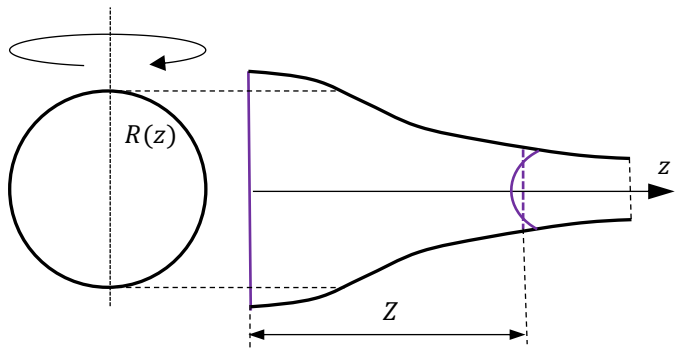

Fig. 3. Schematics of the circular channel model. The broken line is the position of the "flat" tip, $Z$ is the length of the jet, and $R(z)$ is the radius of the channel as a function of the z-coordinate.

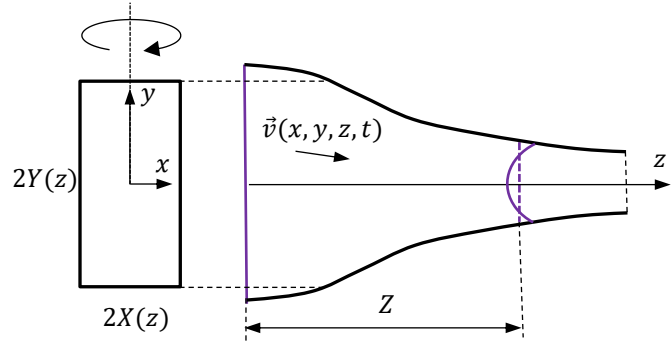

Fig. 4. Schematics of the rectangular channel model. The broken line is the position of the "flat" tip; $Z$ is the length of the jet; and $2X(z)$ and $2Y(z)$ are the width and the length of the cross-section of the channel as a function of the $z$-coordinate.

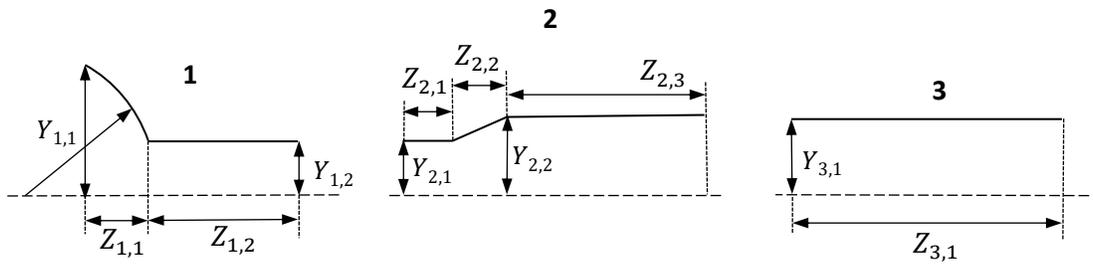

Fig. 5. The two geometries of three-channel Y-shape junction models, not to scale; all dimensions are in micrometers.
Channel 1: $Y_{1,1} = 50$, $Y_{1,2} = 30$, $Z_{1,1} = 300$, $Z_{1,2} = 1200$, $X_1 = 20$ is the width of the channel.
Channel 2: (a) $Y_{2,1} = 15$, (b) $Y_{2,1} = 3$, $Y_{2,2} = 20$, $Z_{2,1} = 200$, $Z_{2,2} = 10000$, $X_2 = 14$ is the width of the channel.
Channel 3: (a) $Y_{3,1} = 15$, (b) $Y_{3,1} = 27$, $Z_{3,1} = 1000$, $X_3 = 20$ is the width of the channel.

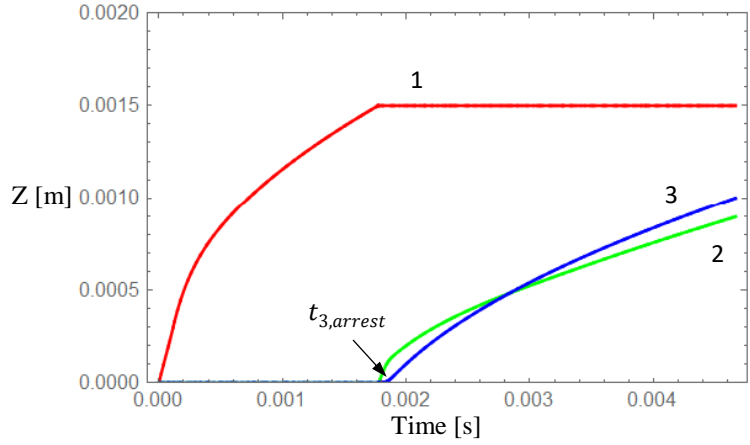

Fig. 6. Simulation of the process of filling of the three-channel Y-shape junction; the full model with $P_0 = 0$. The numbers represent the channel's numbers in Fig. 2; the geometries of channels shown in Fig. 5 case (a); $t_{3,asrrest}$ is the duration of the flow arrest in channel 3. Fig 7 shows the magnified areas of filling of channels 2 and 3.

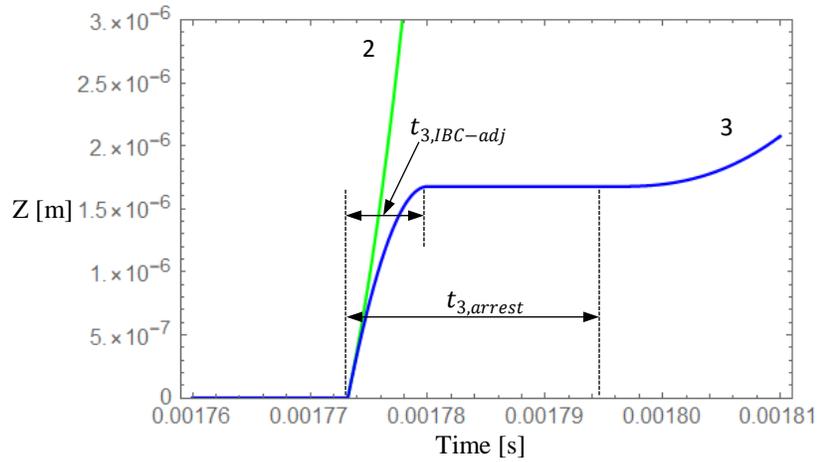

Fig. 7. The filling of channels 2 and 3, the full model, geometry (a) in Fig. 5, and $P_0 = 0$; $t_{3,arrest}$ is the duration of flow arrest in channel 3; $t_{3,IBS-adj}$ is the time for the flow in channel 3 to change from its initial boundary condition value to its plateau value.

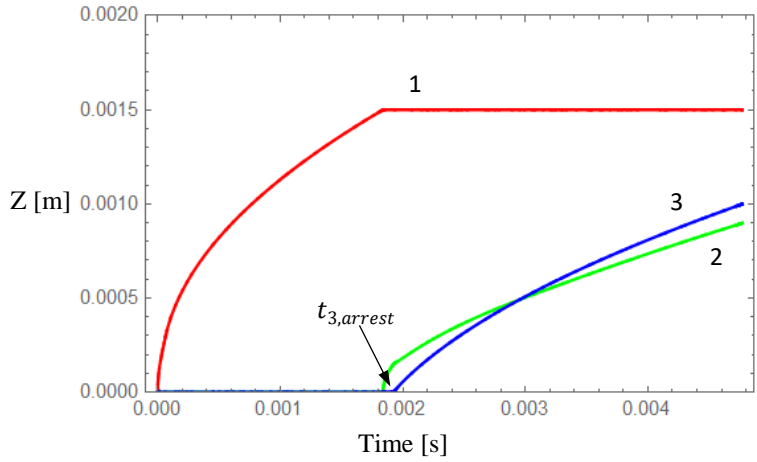

Fig. 8. Simulation of the process of filling of the three-channel Y-shape junction; the reduced model with $P_0 = 0$. The numbers represent the channel number in Fig. 2; the geometries of channels shown in Fig. 5 case (a); $t_{3,asrrest}$ is the duration of the flow arrest in channel 3. Fig. 9 shows the magnified areas of filling of channels 2 and 3.

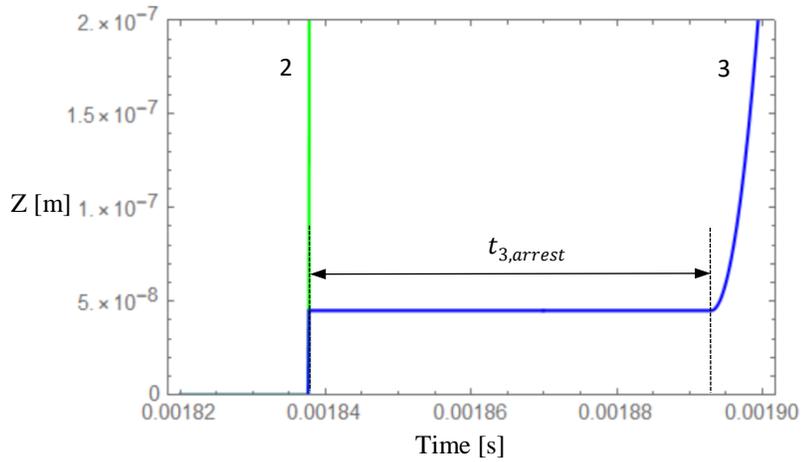

Fig. 9. The filling of channels 2 and 3, the reduced model, geometry (a) in Fig. 5, and $P_0 = 0$; $t_{3,arrest}$ is the duration of the flow arrest in channels 3.

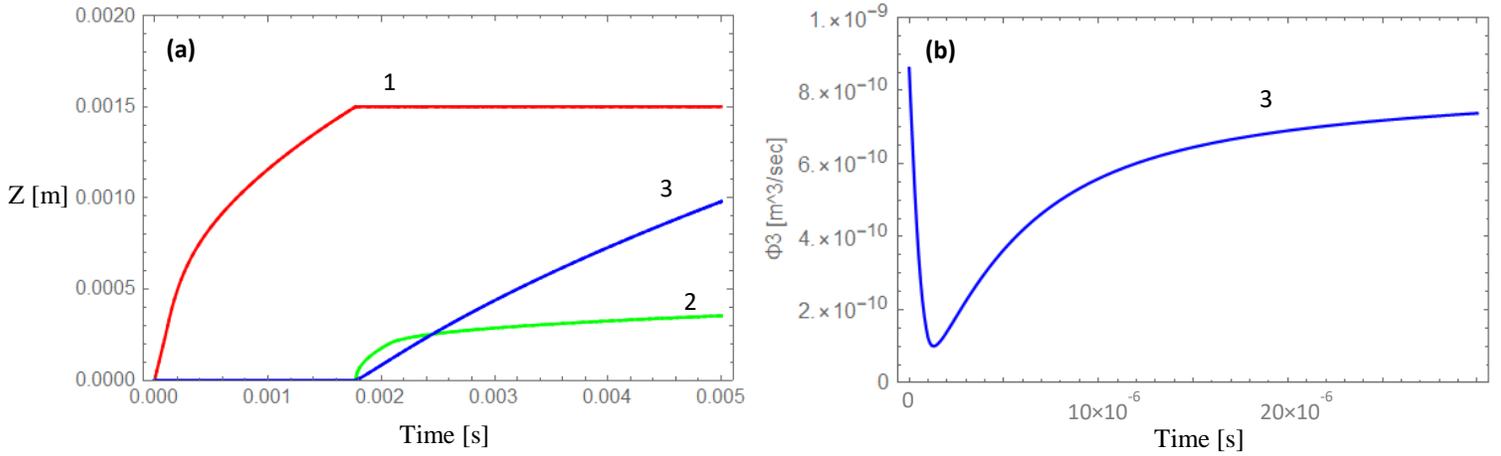

Fig. 10. Simulation of the process of filling of the three-channel Y-shape junction; the full model with $P_0 = 0$. The numbers represent the channel numbers in Fig. 2; the geometries of channels shown in Fig. 5 case (b). (a) – the filling of channels 1, 2, and 3 vs. time, no arrest flow at the interface of the Y-shape junction. (b) – the flow in channel 3 vs. time; $t = 0$ corresponds to the time when the flow reaches the interface between channels 1, 2, and 3, Fig. 2.

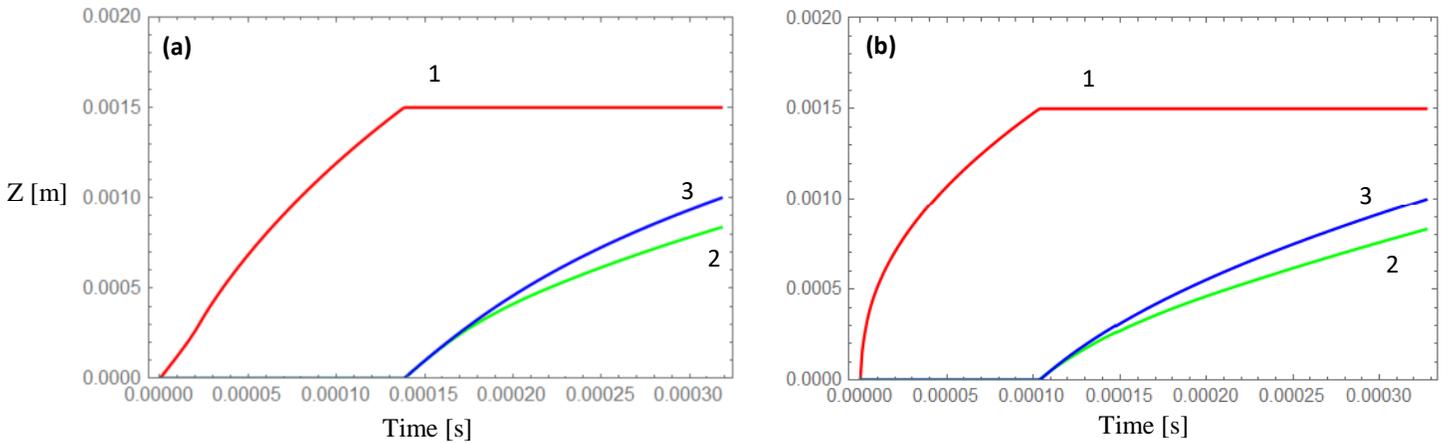

Fig. 11. Simulation of the process of filling of the three-channel Y-shape junction with $P_0 = 10^5 Pa$: (a) - the full model; (b) - the reduced creeping flow model. The numbers represent the channel numbers in Fig. 2; the geometries of channels are shown in Fig. 5, case (a).

39